\documentclass[iop,apjl,tighten]{emulateapj}

\usepackage{natbib}
\slugcomment{To appear in the Astrophysical Journal}

\shorttitle{OXYGEN ABUNDANCES IN NEARBY FGK STARS}
\shortauthors{RAM\'IREZ ET AL.}

\newcommand{\feh}{\mathrm{[Fe/H]}}
\newcommand{\ofe}{\mathrm{[O/Fe]}}
\newcommand{\teff}{T_\mathrm{eff}}
\newcommand{\logg}{\log g}
\newcommand{\afe}{A_\mathrm{Fe}}

\newcommand{\oi}{O\,\textsc{i}}
\newcommand{\fei}{Fe\,\textsc{i}}
\newcommand{\feii}{Fe\,\textsc{ii}}

\newcommand{\kms}{km\,s$^{-1}$}

\begin{document}

\title{OXYGEN ABUNDANCES IN NEARBY FGK STARS AND THE GALACTIC \\ CHEMICAL EVOLUTION OF THE LOCAL DISK AND HALO}

\author{I.~Ram\'irez\altaffilmark{1},
        C.~Allende Prieto\altaffilmark{2,3},
	and
        D.~L. Lambert\altaffilmark{1}
	}

\altaffiltext{1}{McDonald Observatory and Department of Astronomy,
                 University of Texas at Austin,
		 2515 Speedway, Stop C1400, Austin, Texas 78712-1205, USA}
\altaffiltext{2}{Instituto de Astrof\'isica de Canarias,
                 38205, La Laguna, Tenerife, Spain}
\altaffiltext{3}{Departamento de Astrof\'{\i}sica, Universidad de La Laguna,
                 38206, La Laguna, Tenerife, Spain}

\begin{abstract}
Atmospheric parameters and oxygen abundances of 825 nearby FGK stars are derived using high-quality spectra and a non-LTE analysis of the 777\,nm \oi\ triplet lines. We assign a kinematic probability for the stars to be thin-disk ($P_1$), thick-disk ($P_2$), and halo ($P_3$) members. We confirm previous findings of enhanced [O/Fe] in thick-disk ($P_2>0.5$) relative to thin-disk ($P_1>0.5$) stars with $\feh\lesssim-0.2$, as well as a ``knee'' that connects the mean [O/Fe]--[Fe/H] trend of thick-disk stars with that of thin-disk members at $\feh\gtrsim-0.2$. Nevertheless, we find that the kinematic membership criterion fails at separating perfectly the stars in the [O/Fe]--[Fe/H] plane, even when a very restrictive kinematic separation is employed. Stars with ``intermediate'' kinematics ($P_1<0.7$, $P_2<0.7$) do not all populate the region of the [O/Fe]--[Fe/H] plane intermediate between the mean thin-disk and thick-disk trends, but their distribution is not necessarily bimodal. Halo stars ($P_3>0.5$) show a large star-to-star scatter in [O/Fe]--[Fe/H], but most of it is due to stars with Galactocentric rotational velocity $V<-200$\,\kms; halo stars with $V>-200$\,\kms\ follow an [O/Fe]--[Fe/H] relation with almost no star-to-star scatter. Early mergers with satellite galaxies explain most of our observations, but the significant fraction of disk stars with ``ambiguous'' kinematics and abundances suggests that scattering by molecular clouds and radial migration have both played an important role in determining the kinematic and chemical properties of solar neighborhood stars.
\end{abstract}

\keywords{stars: abundances --- stars: atmospheres --- stars: fundamental parameters --- Galaxy: disk --- Galaxy: evolution --- Galaxy: formation}

\section{INTRODUCTION} \label{s:intro}

Stars in the solar neighborhood exhibit a variety of kinematic and elemental abundance properties. Even though a majority of these stars, including the Sun, appear to belong to a common stellar population, namely the Galactic thin disk, a significant number of nearby stars have been associated with the thick disk, whose members tend to be slightly more metal-poor and have enhanced $\alpha$-element abundances (relative to their iron content). Both disk components are rotationally-supported, but the thick disk lags the thin disk, and it has a larger velocity dispersion. A small fraction of nearby stars having very old ages and low metallicities belong to the halo of the Milky Way. Even though restricted to a small region of the Galaxy, disentangling these local populations and studying in detail their properties allow us to investigate the formation and evolution of the Milky Way galaxy.

The obvious advantage of studying nearby stars is the fact that their analysis can be based on very high quality astrometric, photometric, and spectroscopic data. Unfortunately, currently available observational resources allow us only to perform restricted surveys of nearby stars (distances closer than about 250\,pc) at high spectral resolution ($R=\lambda/\Delta\lambda\gtrsim20\,000$), implying severe sample biases and uncertain selection functions. The latter prevent a straightforward comparison of observed chemical abundance patterns with models of Galactic evolution. This observational deficiency will be resolved in the foreseeable future with ongoing surveys such as the Apache Point Observatory Galactic Evolution Experiment \cite[APOGEE, e.g.,][]{allende08:apogee,majewski10,eisenstein11}, which is obtaining $R\simeq20\,000$ infrared spectra of about $10^5$ giant stars, the Gaia-ESO Public Spectroscopic Survey \citep{gilmore12}, whose goal is to acquire visible $R\gtrsim15\,000$ spectra of a similar number of carefully selected dwarf and giant stars in the Milky Way, and planned surveys such as the HERMES project \cite[High Efficiency and Resolution Multi-Element Spectrograph for the Anglo-Australian Telescope; e.g.,][]{freeman10}, which will obtain $R\simeq30\,000$ visible spectra of about $10^6$ stars.

Analyses of nearly complete, unbiased data sets for nearby stars such as the Geneva-Copenhagen Survey \cite[GCS, e.g.,][]{nordstrom04,holmberg07,casagrande11} and low resolution spectroscopic studies of large samples ($>350\,000$) of faint and more distant stars such as the Sloan Extension for Galactic Understanding and Exploration \cite[SEGUE/SDSS, e.g.,][]{yanni09,dejong10,lee11,cheng12a,cheng12b,schlesinger12} are improving significantly our knowledge of Galactic structure, dynamics, and chemistry, thus providing strong constraints to Galaxy formation and evolution scenarios. More details can in principle be seen in higher quality data, implying that it is still helpful at this moment to investigate chemical abundance patterns of few to several hundreds of nearby stars using high spectral resolution, high signal-to-noise ratio data.

Since pioneering work by, e.g., \cite{gilmore83} and \cite{soubiran93}, the picture of a Galactic disk composed of a kinematically cold thin disk and a kinematically warm thick disk, as discrete populations, has been generally accepted \cite[but see, e.g.,][]{norris91,norris99}.\footnote{Disks are often referred to as the kinematically cold components of galaxies, in contrast to the hot halo and (classical) bulge components. In this paper, we refer to the thick disk as the {\it warm} sub-structure of the cold disk, to differentiate it kinematically from the {\it colder} thin disk.} Chemical abundance patterns of kinematically selected samples of solar neighborhood stars appear to support this idea \citep[e.g.,][]{fuhrmann98,bensby05,reddy06,ramirez07}, as well as the analysis of stars in low to medium spectral resolution surveys of much larger volumes such as the Radial Velocity Experiment \cite[RAVE, e.g.,][]{veltz08} and SEGUE/SDSS \cite[e.g.,][]{lee11}. In all these works, as mentioned before, thick-disk stars are found to be generally older and statistically more metal-poor than thin-disk members \cite[e.g.,][]{fuhrmann98,gratton00,bensby05,reddy06,allende06}. The chemical abundances suggest an enhancement of $\alpha$-elements relative to iron for thick-disk stars, although the amount of enhancement depends on the stellar metallicity \cite[e.g.,][]{bensby05,reddy06}. In fact, kinematically-selected thin- and thick-disk stars seem to have about the same $\alpha$-element to iron abundance ratios at $\feh\gtrsim0$.\footnote{In this work we use the standard definitions: $\mathrm{[X/Y]}=\log(N_\mathrm{X}/N_\mathrm{Y})-\log(N_\mathrm{X}/N_\mathrm{Y})_\odot$, and $A_\mathrm{X}=\log(N_\mathrm{X}/N_\mathrm{H})+12$, where $N_\mathrm{X}$ is the number density of nuclei of the element X.}

In the case of local, nearby stars, sample biases can be problematic when interpreting results obtained from the analysis of high quality spectroscopic data. The solar neighborhood is dominated by thin-disk stars. Thus, when thick-disk star samples are constructed, a strong kinematic criterion is often applied to avoid thin-disk ``contamination,'' leading to thick-disk star samples that may be more representative of some extreme kinematic limit of the thick disk distribution rather than the average thick disk. In most cases, an equivalently strong kinematic criterion is applied to construct a comparison sample of thin-disk stars, thus removing anything that could have an intermediate behavior, either in the kinematics or in the chemical abundances. Studying in detail (i.e., with high resolution, high signal-to-noise ratio spectra) objects that have so far been arbitrarily removed from thin/thick disk chemical abundance trends could help us draw a more complete picture of Galactic chemical evolution. In fact, the work by \cite{fuhrmann08}, who uses the magnesium abundance as indicator of $\alpha$-element enhancement, has already hinted at the importance of ``transition'' objects in the Galactic disk \cite[see also][]{fuhrmann98,fuhrmann04}.

Historically, theories for the formation and evolution of the Galactic disk have been divided into two generic groups, namely bottom-up and top-down models \cite[e.g.,][]{majewski93,freeman02}. In the former, a thick disk is formed by secular processes internal to the Galaxy, while in the latter, mergers with satellite galaxies are usually called upon to explain the existence of the thick disk. Until recently, the different merger scenarios seemed to be the only ones capable of reproducing in detail most of the data observed in the solar neighborhood \cite[e.g.,][]{abadi03,brook05}. However, bottom-up models in which radial mixing is an important component of the process of Galactic disk formation have been shown to explain these data as successfully as the merger models, and in some cases even better \cite[e.g.,][]{schonrich09,loebman11,bird12}.  Nevertheless, recent results from the SEGUE/SDSS collaboration show that these models may have difficulties when dealing with observations of the Galaxy on a larger-scale \cite[e.g.,][]{schlesinger12}. Also, radial mixing alone cannot account for the existence of counter-rotating thick disks, which have been observed in some external galaxies \cite[e.g.,][]{yoachim05}. Moreover, \cite{minchev12} have used $N$-body simulations to argue that the thickening of the Galactic disk may not be attributed to radial mixing. Despite all observational and theoretical efforts, we do not yet have a fully consistent picture for the formation and evolution of the Milky Way's disk.

\medskip

In this work, we measure oxygen abundances of a large number of nearby FGK stars, mainly dwarfs and subgiants, to provide additional clues to the formation and evolution of the Galactic disk, and to a lesser extent, that of the Milky Way's halo. This work is an extension of a previous study in which we presented oxygen abundances of 523 nearby stars \citep[][hereafter R07]{ramirez07}. We have increased our sample size by more than 300 objects and have improved the determination of stellar parameters and chemical abundances, as described in Sections~\ref{s:sample} and \ref{s:parameters}. The oxygen abundance patterns derived from our data are presented in Section~\ref{s:trends}, along with a discussion relevant to the chemical evolution of the Galaxy. Our conclusions and final remarks are given in Section~\ref{s:conclusions}.

\section{SAMPLE AND BASIC DATA} \label{s:sample}

\subsection{Spectroscopic Data}

We have analyzed 897 spectra of 825 stars, in addition to 10 solar (day-sky and asteroid) spectra. Most of these spectra were taken by us, but we also used data from public archives. Multiple spectra for a number of stars are available from more than one source. Instead of combining them, which would not be trivial given the differences in spectral resolution and the particular way in which the continuum normalization was made, in addition to differences in wavelength coverage, we analyzed each spectrum independently and averaged the final results for the stellar parameters and elemental abundances. All spectra used have high resolution ($R\gtrsim45,000$) and high signal-to-noise ratio ($S/N\gtrsim100$). This minimizes the impact of line blends and allows an accurate determination of local continua, which are important to reduce the observational errors.

Our spectra have been acquired using four instrument/telescope combinations, which are listed here, along with their corresponding references, in order of importance, as quantified by the number of spectra taken from each source: TS2/McD (R.\ G.\ Tull Coud\'e spectrograph, 2.7\,m Telescope at McDonald Observatory; \citealt{tull95}), HRS/HET (High Resolution Spectrograph, 9.2\,m Hobby-Eberly Telescope; \citealt{tull98}), UVES/VLT (UV-Visual Echelle Spectrograph, 8\,m Very Large Telescope; \citealt{dekker00}), and FEROS/ESO (Fiber-feb Extended Range Optical Spectrograph, ESO 1.52-m Telescope; \citealt{kaufer98}). Details on the spectroscopic data reduction can be found in the references given below.

\begin{deluxetable}{lcrcr}
\tablecaption{Sources of Spectroscopic Data}
\tablehead{\colhead{Source\tablerefs{\citet[][B03]{bagnulo03}, \citet[][R03]{reddy03}, \citet[][R06]{reddy06}, \citet[][AP04]{allende04:s4n}, \citet[][R07]{ramirez07}, \citet[][AP08]{allende08a}, \citet[][R09]{ramirez09}, This Work (TW).}} & \colhead{Facility/Instrument} & \colhead{$R$} & \colhead{$S/N$} & \colhead{Spectra}}
\tablewidth{0pc}
\startdata
B03       & VLT/UVES          & 80\,000                   & 300--500  & 55        \\
R03/R06   & McD/TS2	      & 60\,000 		  & 100--400  & 369       \\
AP04\tablenotemark{1}      & McD/TS2	      & 60\,000 		  & 150--600  & 106       \\
\nodata & ESO/FEROS         & 45\,000                   & \nodata & \nodata \\
R07       & HET/HRS           & 120\,000                  & 200--300  & 52        \\
\nodata & McD/TS2	    & 60\,000 		        & 150--250  & 26        \\
AP08      & HET/HRS           & 120\,000                  & 100--200  & 86        \\
R09       & McD/TS2	      & 60\,000 		  & 150--300  & 69        \\
TW        & McD/TS2	      & 60\,000 		  & 100--300  & 144       

\enddata
\tablenotetext{1}{The number of spectra from AP04 correspond to the combined McD/TS2 and ESO/FEROS data set.}
\label{t:spectra}
\end{deluxetable}

The sources of our spectra are listed in Table~\ref{t:spectra}. Spectra from \cite{reddy03,reddy06}, \cite{allende04:s4n}, and \cite{bagnulo03} were already analyzed in R07, where we also included 78 stars from our own observations. For this work, we add the observations of solar analog stars by \cite{allende08a} and \cite{ramirez09}. All these spectra have been re-analyzed in this work. In addition, we analyze new data we have obtained for 144 stars with the Tull spectrograph at McDonald Observatory. Most of these stars were selected to increase the number of stars with kinematic properties intermediate between those of the Galactic thin and thick disks. The reduction of these new data followed the exact same procedure described in R07.

\begin{figure}
\includegraphics[bb=100 370 510 910,width=9.0cm]
{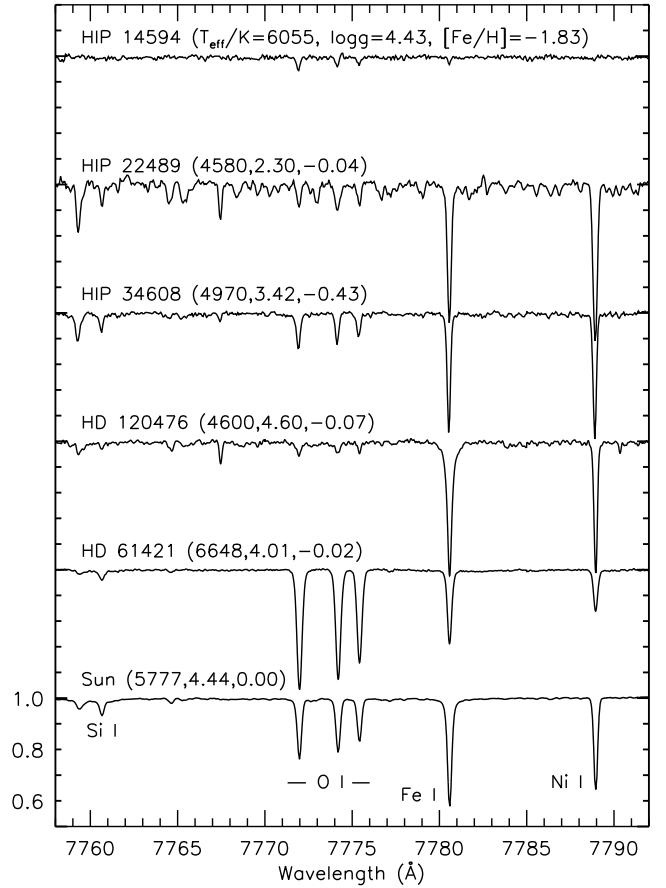}
\caption{Sample spectra used in this work. Only a very small wavelength window centered at the \oi\ 777\,nm triplet is shown. The numbers in parenthesis are the atmospheric parameters $\teff$ (in K), $\logg$, and $\feh$ of each star.}
\label{f:triplet_sample}
\end{figure}

A few sample spectra are plotted in Figure~\ref{f:triplet_sample}. The strongest features seen in the solar spectrum are identified. Besides showing the high quality of the data used, this plot roughly traces the behavior of the 777\,nm \oi\ triplet lines, which we use as oxygen abundance indicators, across the region of stellar atmospheric parameter space covered by our sample.

In all cases shown in Figure~\ref{f:triplet_sample}, the 777\,nm \oi\ triplet is clearly identified and easily measured. At the spectral resolution of most of our data, line asymmetries due to surface convection are negligible, which allows us to use Gaussian or Voigt profile fits to accurately quantify line strengths. Line blending of the triplet is negligible in most FGK dwarf and subgiant stars, and it is important only for the coolest stars of our sample. Contamination by CN features can be seen in the spectrum of the cool giant star HIP\,22489, as well as clear evidence for a blend in the middle line of the triplet, which is expected to have an intensity intermediate between those of the red and blue components. This blend is probably due to an \fei\ line \cite[e.g.,][]{takeda98,schuler06:giant-dwarf}. The middle line of the triplet was excluded from our cool K-dwarf and cool giant star analysis.

It is important to note that there are no strong telluric features contaminating the wavelength region shown in Figure~\ref{f:triplet_sample}. They would appear obvious as narrow spectral lines located at different wavelengths in each spectrum, given that these sample spectra are corrected for the stellar radial velocities. Visual inspection of several telluric standard star observations confirms this statement. Moreover, the continuum normalization of this spectral window is simple since no extremely strong lines are present and relatively few spectral lines are visible at all, except in the case of the cool giant star HIP\,22489.

\subsection{Photometry}

Visual magnitudes $V$ and $(B-V)$ colors were extracted mainly from the General Catalogue of Photometric Data \citep[GCPD,][]{mermilliod97}.\footnote{Available online at http://obswww.unige.ch/gcpd/gcpd.html} For the few stars without Johnson's photometry in the GCPD, we searched the {\it Hipparcos} catalog \citep{perryman97} for $V$ and $(B-V)$ values observed from the ground. These are values compiled from the literature by the {\it Hipparcos} team and not values inferred from transformation equations using Tycho photometry, which are not so accurate. Cousins $RI$ photometry was also extracted from the GCPD, where available. For stars fainter than $V\simeq6$, we used $JHK_s$ photometry from the Two Micron All Sky Survey \citep[2MASS,][]{cutri03}. For brighter stars, 2MASS data are less reliable due to saturation. Str\"omgren $(b-y)$ colors were also used, where available, as listed in the GCPD and the Geneva-Copenhagen Survey \citep{nordstrom04,casagrande11}.

\subsection{Distance, Proper Motion, and Radial Velocity}

Most of our sample stars are included in the {\it Hipparcos} catalog. We used the parallaxes and proper motions from the new reduction by \cite{vanleeuwen07}. Stellar distances were derived from these trigonometric parallaxes. For HD\,144070, which is not included in the {\it Hipparcos} catalog, we used the ground-based measurement listed in the \cite{vanaltena95} compilation. The stars HIP\,7751 and HIP\,55288 are each known to be in wide visual binary systems. Their secondaries, for which we also have spectroscopic data and have therefore been analyzed, are fainter and not listed in the {\it Hipparcos} catalog. In these cases we adopted the parallaxes of the primaries for the secondaries. Our spectroscopic analysis confirmed the true binary nature of these two pairs; the radial velocities and elemental abundances we infer are in agreement for the two stars in each pair.

Radial velocities were adopted from published catalogs and our own measurements. In R07, we provided radial velocities for a large fraction of our sample stars. For these objects, the radial velocities adopted here are from that paper. For the other stars, we used the average of values from the following sources: \cite{barbier-brossat94}, \cite{duflot95}, the Geneva-Copenhagen Survey \citep{nordstrom04}, \cite{ramirez09}, \cite{jenkins11}, \cite{latham02}, and our own measurements. Other sources for few stars, which were not found in any of the works listed above, are \cite{famaey05}, \cite{evans69}, and \cite{wilson53}.

Most stars have radial velocities with uncertainties under 1\,\kms. A few are known spectroscopic binaries where the spectrum of the secondary is not detected, but important radial velocity variations are measured. For these objects we searched for orbital solutions in the  \cite{latham02} survey and adopted their mean system velocity as the radial velocity of the star. A few of them do not have orbital solutions published and we therefore adopted a large error corresponding to the maximum difference in radial velocity observed.

\subsection{Galactic Space Velocities and Kinematic Membership Criteria} \label{s:kinematics}

With the radial velocities and proper motions we derived heliocentric Galactic space velocities $U,V,W$ following the recipe by \cite{johnson87}. Errors in the input data were propagated as suggested by them. We note, however, that this recipe does not take into account the correlations between uncertainties in the astrometric quantities that are provided in the {\it Hipparcos} catalog. An improved error treatment is given in the {\it Hipparcos} catalog \citep{perryman97} or in \cite{allende04:s4n}. For stars that do not have an {\it Hipparcos} parallax we adopted the $U,V,W$ values from the Geneva Copenhagen Survey \cite[e.g.,][]{casagrande11} and assumed conservative $U,V,W$ errors of 2.5\,\kms, as these values are not given for individual stars in the GCS.

Membership probabilities (i.e., the probabilities of a star to be a thin disk, thick disk, and halo member) were computed as in R07 (their Section~3.3; see also \citealt{mishenina04}). In summary, stars of a given population are assumed to have a Gaussian distribution in their Galactic space velocities, with mean $U,V,W$ values and velocity dispersions ($\sigma$) adopted from \cite{soubiran03} for the thin and thick disks, and from \cite{chiba00} for the halo.\footnote{Thin disk: ($V_1,\sigma U_1,\sigma V_1,\sigma W_1)=(-12,39,20,20)$\,\kms, thick disk: ($V_2,\sigma U_2,\sigma V_2,\sigma W_2)=(-51,63,39,39)$\,\kms, and halo: ($V_3,\sigma U_3,\sigma V_3,\sigma W_3)=(-199,141,106,94)$\,\kms.} The probability that a star with Galactic space velocity components $U,V,W$ belongs to the thin disk ($P_1$), thick disk ($P_2$), or halo ($P_3$) is thus given by:
\begin{eqnarray}
P_i & = & \frac{c_i}{(2\pi)^{3/2}\sigma_{U_i}\sigma_{V_i}\sigma_{W_i}} \times  \\
    &   & \exp\left\{-0.5\left[\frac{U^2}{\sigma^2_{U_i}}+\frac{(V-V_i)^2}{\sigma^2_{V_i}} +\frac{W^2}{\sigma^2_{W_i}}\right]\right\} , \nonumber
\end{eqnarray}
where
\begin{equation}
c_i=\frac{p_i}{\sum_{i=1}^3p_i(P_i/c_i)}
\end{equation}
is a normalization constant which ensures that $\sum P_i=1$. The $p_i$ values are the relative number densities of thin-disk, thick-disk, and halo stars in the solar neighborhood, for which we adopt $p_1=0.90$, $p_2=0.08$, and $p_3=0.02$.

In this work, we associate stars with $P_1>0.5$ with the thin disk (i.e., those with 50\,\% or greater probability of being thin-disk members), $P_2>0.5$ with the thick disk, and $P_3>0.5$ with the halo. In order to compare our results with other works, in certain parts of this paper we will adopt, in addition to the former conditions, a ``strong'' kinematic criterion of $P_1/P_2>10$ for the thin disk and $P_2/P_1>10$ for the thick disk \cite[cf.][]{bensby04,bensby05}. Note, however, that the latter criterion implies that a number of stars will be rejected from the analysis, which, as we have argued in Section~\ref{s:intro}, is something that could produce biased results.

\medskip

\begin{deluxetable*}{lcrrrrrccc}
\tablecaption{Basic and Kinematic Data\tablenotemark{1}}
\tablehead{\colhead{Star} & \colhead{$V_\mathrm{mag}$} & \colhead{$\pi$ (mas)} & \colhead{$V_r$ (\kms)} & \colhead{$U$ (\kms)} & \colhead{$V$ (\kms)} & \colhead{$W$ (\kms)} & \colhead{$P_1$} & \colhead{$P_2$} & \colhead{$P_3$}\tablewidth{0pc}}
\tabletypesize{\scriptsize}
\tablewidth{0pc}
\startdata
  HD 32071 &   8.93 &        \nodata &$   8.3\pm  0.5$&$ -24.0\pm  2.5$&$ -25.0\pm  2.5$&$  56.0\pm  2.5$&  0.77 &  0.22 &  0.00 \\
  HD 59490 &   8.68 &        \nodata &$  91.0\pm  0.5$&$ -83.0\pm  2.5$&$ -49.0\pm  2.5$&$ -17.0\pm  2.5$&  0.70 &  0.30 &  0.00 \\
  HD 67163 &   8.05 &        \nodata &$  65.3\pm  0.5$&$ -16.0\pm  2.5$&$ -64.0\pm  2.5$&$  -4.0\pm  2.5$&  0.70 &  0.30 &  0.00 \\
  HD 82960 &   8.53 &        \nodata &$  46.0\pm  0.5$&$  28.0\pm  2.5$&$ -58.0\pm  2.5$&$   5.0\pm  2.5$&  0.80 &  0.19 &  0.00 \\
 HD 130047 &   8.56 &        \nodata &$  15.9\pm  0.5$&$  41.0\pm  2.5$&$ -63.0\pm  2.5$&$ -21.0\pm  2.5$&  0.57 &  0.43 &  0.00 \\
 HD 144070 &   4.84 &   $43.0\pm4.2$ &$ -32.6\pm  1.1$&$ -30.0\pm  2.5$&$  -6.0\pm  2.5$&$ -14.0\pm  2.5$&  0.99 &  0.01 &  0.00 \\
 HD 170058 &   8.42 &        \nodata &$ -21.2\pm  0.5$&$   3.0\pm  2.5$&$ -49.0\pm  2.5$&$ -28.0\pm  2.5$&  0.86 &  0.14 &  0.00 \\
 HD 171029 &   8.24 &        \nodata &$ -37.8\pm  0.5$&$ -22.0\pm  2.5$&$ -65.0\pm  2.5$&$  -7.0\pm  2.5$&  0.65 &  0.34 &  0.00 \\
 HD 183490 &   8.22 &        \nodata &$ -66.1\pm  0.7$&$  -5.0\pm  2.5$&$ -71.0\pm  2.5$&$   0.0\pm  2.5$&  0.50 &  0.49 &  0.00 \\
 HD 213746 &   8.55 &        \nodata &$ -59.1\pm  1.9$&$  -2.0\pm  2.5$&$ -71.0\pm  2.5$&$ -15.0\pm  2.5$&  0.45 &  0.54 &  0.01 \\
 HD 223723 &   8.60 &        \nodata &$   4.3\pm  0.5$&$ -85.0\pm  2.5$&$ -35.0\pm  2.5$&$ -26.0\pm  2.5$&  0.83 &  0.17 &  0.00 \\
   HIP 171 &   5.74 & $82.17\pm2.23$ &$ -36.9\pm  0.5$&$  -8.0\pm  0.7$&$ -73.1\pm  1.7$&$ -31.1\pm  2.0$&  0.23 &  0.76 &  0.01 \\
   HIP 348 &   8.62 & $17.43\pm0.90$ &$  18.1\pm  0.5$&$ -94.0\pm  6.1$&$ -12.1\pm  1.6$&$ -12.2\pm  0.4$&  0.94 &  0.06 &  0.00 \\
   HIP 394 &   6.11 & $25.52\pm3.28$ &$   4.6\pm  0.5$&$-124.8\pm 19.4$&$ -51.3\pm  8.7$&$ -10.9\pm  1.5$&  0.27 &  0.71 &  0.02 \\
   HIP 475 &   8.22 & $19.01\pm0.82$ &$ -28.4\pm  0.5$&$   6.8\pm  0.4$&$ -31.4\pm  0.6$&$ -36.7\pm  2.0$&  0.93 &  0.07 &  0.00 \\
   HIP 493 &   7.45 & $26.93\pm0.56$ &$ -45.6\pm  0.5$&$  45.8\pm  1.1$&$ -34.2\pm  0.4$&$  17.0\pm  0.6$&  0.95 &  0.05 &  0.00 \\
   HIP 522 &   5.70 & $38.89\pm0.37$ &$   5.3\pm  0.5$&$ -57.4\pm  1.0$&$ -37.5\pm  0.6$&$ -15.3\pm  0.5$&  0.93 &  0.07 &  0.00 \\
   HIP 530 &   8.35 &  $9.38\pm0.71$ &$ -31.8\pm  0.5$&$ -28.2\pm  4.5$&$ -48.5\pm  2.4$&$   2.8\pm  0.9$&  0.92 &  0.08 &  0.00 \\
   HIP 544 &   6.10 & $73.15\pm0.56$ &$  -6.9\pm  0.5$&$ -15.0\pm  0.2$&$ -21.6\pm  0.4$&$ -10.1\pm  0.3$&  0.99 &  0.01 &  0.00 \\
   HIP 656 &   8.15 &  $9.14\pm0.98$ &$ -42.5\pm  0.5$&$  -5.8\pm  2.5$&$ -55.7\pm  1.9$&$ -29.0\pm  3.6$&  0.74 &  0.25 &  0.00 \\

\ldots & \ldots & \ldots & \ldots & \ldots & \ldots & \ldots & \ldots & \ldots & \ldots 
\enddata
\tablenotetext{1}{Table~\ref{t:basic} is published in its entirety in the electronic edition of the Astrophysical Journal. A portion is shown here for guidance regarding its form and content.}
\label{t:basic}
\end{deluxetable*}

\begin{figure}
\includegraphics[bb=60 360 620 575,width=8.8cm]{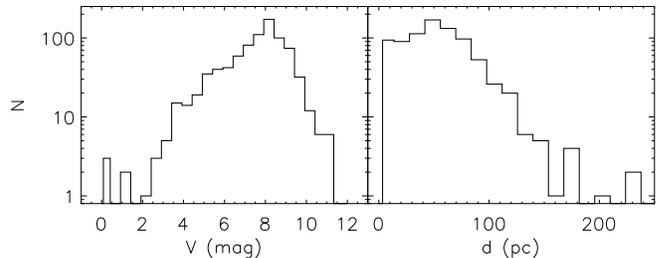}
\caption{Distribution of visual magnitude and distance for our sample stars. Most of them are brighter than $V=10$\,mag and closer than $d=200$\,pc.}
\label{f:vmagd}
\end{figure}

\begin{figure}
\includegraphics[bb=85 375 510 690,width=9cm]{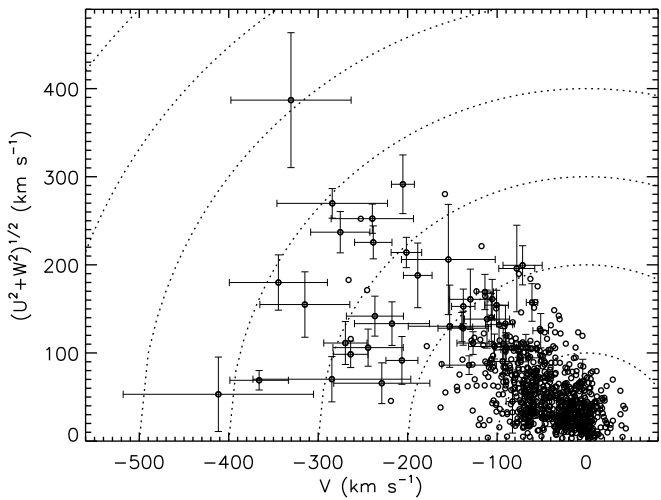}
\caption{Toomre diagram for our sample stars. Error bars are shown only for stars with large velocity errors ($>20$\,\kms\ in the total Galactic space speed). Dotted lines show curves of constant speed, i.e., $(U^2+V^2+W^2)^{1/2}=\mathrm{const.}$}
\label{f:toomre}
\end{figure}

We provide in Table~\ref{t:basic} visual magnitudes, trigonometric parallaxes, radial velocities, Galactic space velocities, and membership probabilities along with their formal errors, for all our sample stars. Figure~\ref{f:vmagd} shows the distribution of apparent brightness and distance for these objects. A Toomre diagram is shown in Figure~\ref{f:toomre}.

\section{STELLAR PARAMETERS AND ABUNDANCES} \label{s:parameters}

\begin{figure}
\includegraphics[bb=80 370 450 658,width=9.0cm]{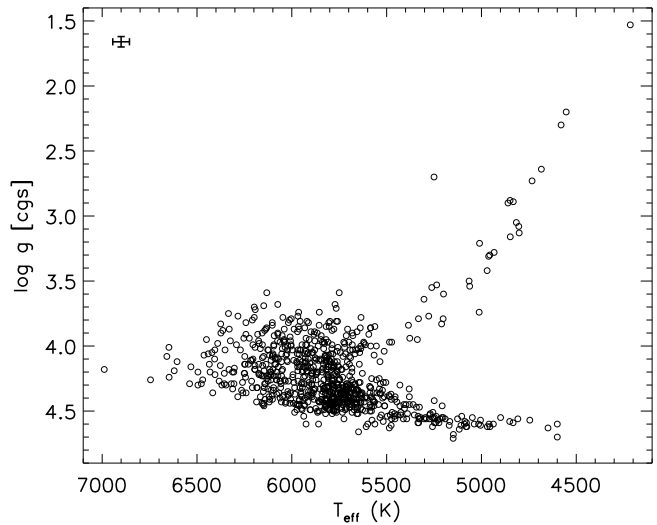}
\caption{HR diagram of our sample. The error bar on the top left shows the average size of our internal uncertainties, excluding systematic errors.}
\label{f:hr}
\end{figure}

The majority of our sample objects are dwarf, turn-off, and subgiant stars with $\feh\gtrsim-1.5$ and a reliable trigonometric parallax measurement. The stellar parameters for these stars, hereafter referred to as the ``main'' sample, were determined as described in Sects.~\ref{s:teff} to \ref{s:iron}. For the other stars, we used alternative methods, as explained in Section~\ref{s:special}. Stellar parameter determination is an iterative procedure because the end products $\teff,\logg,\feh$ are inter-dependent. The results presented below were obtained after several iterations, which ensures that, for each star, the three parameters are self-consistent. An HR diagram of the full sample, using our final atmospheric parameters, is shown in Figure~\ref{f:hr}.

\subsection{Effective Temperature} \label{s:teff}

Using as many as available of the $(B-V)$, $(V-R_\mathrm{C})$, $(V-I_\mathrm{C})$, $(R_\mathrm{C}-I_\mathrm{C})$, $(V-J)$, $(V-H)$, $(V-K_s)$, $(J-K_s)$, and $(b-y)$ colors, we derived the star's effective temperatures from the metallicity-dependent color-$\teff$ calibrations by \cite{casagrande10}. Since one $\teff$ value is obtained from each color, we computed a weighted mean $\teff$ value and error for each star. The weights were computed using the photometric errors and the mean accuracy of each color calibration, as given by \cite{casagrande10}. Since the color-calibrations used depend on $\feh$, the error in $\feh$ that we obtain in Section~\ref{s:iron} was also taken into account when estimating the error in $\teff$. Most of our sample stars have at least five colors available, and their $\teff$ values have an average 1\,$\sigma$ color-to-color scatter of 48\,K.

Interstellar reddening is important only for stars more distant than about 100\,pc, particularly if they are located close to the Galactic plane. As shown by, e.g., \cite{lallement03}, the Sun is located inside a ``Local Bubble'' of low interstellar gas density which appears to be dust free. This region has a radius of about 75\,pc, but is far from spherically symmetric \cite[e.g.,][]{leroy99,welsh10}. We employed several interstellar reddening maps to derive $E(B-V)$ values for stars more distant than 75\,pc \cite[as described in detail in][their Section~3.1]{ramirez05a}. For stars closer than 75\,pc we assumed $E(B-V)=0$.

The temperature-color calibrations by \cite{casagrande10} are based on effective temperatures obtained in the most recent implementation of the so-called infrared flux method \citep[IRFM, e.g.,][]{blackwell77,blackwell79}. The zero point of the \cite{casagrande10} $\teff$ scale has been carefully calibrated and successfully tested using samples of well-known stars and precise spectrophotometric data. It surpasses previous implementations of the IRFM, in particular those by \cite{alonso96} and \cite{ramirez05a}. The latter was used in R07, and this constitutes one of the major improvements made to our work, particularly considering the fact that the zero point of the IRFM $\teff$ scale was shifted by about 100\,K in the more recent work by \cite{casagrande10}. This $\teff$ offset translates into important changes to the $\feh$ and $\ofe$ elemental abundance scales, as will be discussed later.

\subsection{Surface Gravity, Mass, and Age} \label{s:logg}

Surface gravities $(\logg)$, stellar masses ($M$), and ages ($\tau$) were computed by comparing the location of our sample stars in the HR diagram ($\teff$ vs.~$M_V$) with theoretical predictions based on stellar evolution calculations. The absolute magnitudes were obtained from the stars' apparent magnitudes and trigonometric parallaxes. We used the Yonsei-Yale isochrones, which take into account $\alpha$-element enhancement at low metallicities \citep{yi01,yi03,kim02}.

In our age determination algorithm, each isochrone point is represented by a set of parameters $(x,t,m,f)$, where $x$ could be the surface gravity, mass, or age; $t$ is the effective temperature, $m$ the absolute magnitude, and $f$ the iron abundance. To obtain the most likely parameter $x$ given a set of observed quantities ($\teff$, $M_V$, $\feh$) and their associated errors ($\Delta\teff$, $\Delta M_V$, $\Delta\feh$), we use a probabilistic approach \cite[see also, e.g.,][]{reddy03,allende04:s4n,baumann10,chaname12,melendez12}. The probability that a given isochrone point corresponds to the observation is computed as:
\begin{eqnarray}
p(x,t,m,f) & \propto  & \exp [ -(\teff-t)^2 / (\sqrt{2}\Delta\teff)^2 ]\times \nonumber \\
           &          & \exp [ -(M_V-m)^2 / (\sqrt{2}\Delta M_V)^2 ]\times \\
           &          & \exp [ -(\feh-f)^2 / (\sqrt{2}\Delta\feh)^2 ]\ . \nonumber 
\end{eqnarray}
Then, we calculate a probability distribution for the parameter $x$ as follows:
\begin{equation}
P(x)=\sum_{x'=x-\delta x}^{x'=x+\delta x} p(x',t,m,f)\ ,
\end{equation}
where $\delta x$ is a small step for the parameter $x$ while the sum extends also to all values of $t$, $m$, and $f$ within a radius of 3 times the observational errors from the stellar parameters. Including isochrone points beyond these limits has no significant impact on the final result. The $\delta x$ values correspond to a twentieth part (1/20) of the range of $x$ available in each calculation. However, in cases where this $\delta x$ is too small considering the sampling of isochrones, we adopted 0.2\,Gyr and $0.005\,M_\odot$ for the age and mass distributions, respectively. The peak of the $P(x)$ function, which corresponds to the $x$ value of the bin in which $P(x)$ is maximum, is adopted as the most probable surface gravity, mass, or age, while the 68\,\%\ and 96\,\%\ confidence limits are adopted, respectively, as the 1\,$\sigma$ and 2\,$\sigma$ Gaussian-like lower and upper limits.

The isochrone grid used has very fine spacing, in particular in $\feh$, where it is only 0.02\,dex, which was achieved by employing the interpolation routine by \cite{kim02}. This allows us to determine with more precision the shapes of the age probability distributions without having to increase arbitrarily the error bar in $\feh$, as it is sometimes done. The latter could introduce biases in the age determination scheme.

A number of studies have shown that the probabilistic age determination approach that we have adopted is affected by biases that can often overestimate the stellar ages \cite[e.g.,][]{pont04,jorgensen05}. These biases are relevant for works where the absolute values of the ages are important, such as those dealing with the age-metallicity relation. Relative ages or sample chronology (i.e., simply sorting the stars according to their age) are less affected by these biases, yet not completely free from them. In any case, we note that a comparison of our ages with those determined using Bayesian approaches that attempt to minimize those biases, specifically those by \cite{dasilva06} and \cite{casagrande11}, reveals surprisingly good agreement, as shown by \citet[][their Figure~7]{chaname12}, although in part this could also be due to the use of different isochrone sets which may compensate the differences due to the statistical biases.

\subsection{Iron Abundance} \label{s:iron}

We determined $\feh$ values using a standard spectroscopic approach. The 2010 version of the spectrum synthesis code MOOG \citep{sneden73}\footnote{http://www.as.utexas.edu/$\sim$chris/moog.html} and MARCS model atmospheres \citep{gustafsson10} with standard composition (i.e., including $\alpha$-element enhancement at $\feh<0$),\footnote{Downloaded from http://marcs.astro.uu.se. These standard composition MARCS models use scaled solar abundances (those by \citealt{grevesse07}), but with [$\alpha$/Fe] and [O/Fe] abundance ratios increasing linearly from +0.00 to +0.40 as $\feh$ decreases from +0.00 to -1.00. For the $\feh<-1.00$ models, [$\alpha$/Fe] and [O/Fe] are set equal to +0.40.} interpolated linearly to the atmospheric parameters of each star, were used. A large number of neutral (\fei, up to 128) and singly-ionized (\feii, up to 16) iron lines were used for the determination of $\feh$. The version of the spectrum synthesis code and the model atmosphere grid adopted are different from those used in our previous work. In R07, we used the 2002 version of MOOG and Kurucz models with solar-scaled composition, i.e., without $\alpha$-element enhancement.

Equivalent widths were measured using an automated routine capable of identifying blends.\footnote{This routine, ``getew\_xy.pro,'' written in IDL, is available online at http://hebe.as.utexas.edu/stools. Also available at this website is the interpolation routine for MARCS model atmospheres used in this work, ``mmod.pro,'' which was adapted from the code ``kmod.pro'' that works with Kurucz models.} Spectral lines are assumed to have Gaussian shape, which is a poor approximation for strong lines with extended wings. However, we use only relatively weak lines for our iron abundance analysis, thus minimizing this systematic error. Only for very metal-poor stars, the equivalent widths were measured manually using IRAF's task \verb"splot",\footnote{IRAF is distributed by the National Optical Astronomy Observatories, which are operated by the Association of Universities for Research in Astronomy, Inc., under cooperative agreement with the National Science Foundation -- http://iraf.noao.edu} because even small errors in the determination of the local continuum result in a large percent error in the measured equivalent width of very weak lines. For very metal-poor stars we find that the local continuum of weak lines is best determined visually (rather than assuming that it is exactly at 1, which would imply a perfect continuum normalization). Since fewer lines are available for these objects, the higher precision in the equivalent width measurements that can be easily achieved with this approach results in more reliable average elemental abundances.

Average $\feh$ values were obtained from all \fei\ and \feii\ lines analyzed, not as the average of \fei\ and \feii\ abundances measured separately, but giving equal weight to each of the iron lines (both neutral and singly ionized) measured. This procedure is safe because the mean iron abundances derived from \fei\ and \feii\ separately agree reasonably well, as discussed in Section~\ref{s:spectroscopic_equilibrium}. The $\feh$ values used for tracing iron versus oxygen abundance trends in most of Section~\ref{s:trends}, however, are those inferred from the \feii\ lines, because they are less affected by systematic errors on both the spectral line calculations and on the adopted stellar parameters, as will be discussed in Section~\ref{s:fine-tuning}. Errors in $\feh$ were determined by adding in quadrature the standard error of the line-by-line iron abundance scatter and the small changes in $\feh$ due to $\pm1\,\sigma$ variations of $\teff$ and $\logg$.

\subsubsection{Line Selection and Solar Analysis}

Our iron line selection and atomic data adopted are largely from R07. A few additional spectral lines and some minor updates to the atomic data were made following the works by \cite{melendez09:twins} and \cite{asplund09:review}. The former is a high precision elemental abundance study of solar twin stars whereas the latter represents the most complete solar abundance analysis to date. The addition of these lines allowed us to have a better coverage of excitation potential and line strength with very clean features. The pressure broadening damping constants adopted are from \cite{barklem00} and \cite{barklem05}, but for lines without damping constants computed by them, we used the values obtained from Uns\"old's formula, multiplied by a factor of 3. Our adopted iron line list is given in Table~\ref{t:linelist}.

\begin{deluxetable}{ccccr}
\tablecaption{Iron Line List\tablenotemark{1}}
\tablehead{\colhead{$\lambda$ (\AA)} & \colhead{Species} & \colhead{EP (eV)} & \colhead{$\log gf$} & \colhead{$EW_\odot$}}
\tabletypesize{\footnotesize}
\tablewidth{0pc}
\startdata
4630.12 & \fei\  & 2.280 & $-2.52$ & 72.4 \\
4635.85 & \fei\  & 2.850 & $-2.34$ & 54.1 \\
4683.56 & \fei\  & 2.830 & $-2.41$ & 55.9 \\
4690.14 & \fei\  & 3.690 & $-1.61$ & 55.5 \\
4745.80 & \fei\  & 3.650 & $-1.27$ & 78.3 \\
\ldots & \ldots & \ldots & \ldots & \ldots \\
4576.33 & \feii\ & 2.844 & $-2.95$ & 64.4 \\
4620.51 & \feii\ & 2.828 & $-3.21$ & 52.4 \\
4629.34 & \feii\ & 2.810 & $-2.28$ & 97.2 \\
5197.57 & \feii\ & 3.231 & $-2.22$ & 81.7 \\
5234.62 & \feii\ & 3.221 & $-2.18$ & 83.9 \\
\ldots & \ldots & \ldots & \ldots & \ldots 
\enddata
\tablenotetext{1}{Table~\ref{t:linelist} is published in its entirety in the electronic edition of the Astrophysical Journal. A portion is shown here for guidance regarding its form and content.}
\label{t:linelist}
\end{deluxetable}

Ten spectra were available for a solar analysis. Half of them correspond to daysky observations whereas the other half are reflected Sun-light observations of bright asteroids (Ceres, Iris, Pallas, and Vesta). All these spectra are of very high quality and the measured equivalent widths agree very well, with the exception of one daysky spectrum, as discussed below. The average standard deviation of $EW$ values measured for the same line in all solar spectra is 1.8\,m\AA, which corresponds to a standard error for the mean $EW$ of 0.6\,m\AA. Since we used our automated routine to measure the solar $EW$s, this result gives us a rough estimate of our automated $EW$ measurement errors (about 1\,\%).

Only one of our solar spectra revealed a small but non-negligible offset in the measured $EW$s. This daysky spectrum was taken with the HRS/HET at sunset and pointing the telescope to the East, resulting in a large separation angle between the area of the sky observed and the solar position. As shown by \citet[][see also Section~2 in \citealt{ramirez11}]{gray00}, scattered light in the Earth's atmosphere can affect significantly the shapes and also the strengths of spectral lines observed under these conditions (i.e., at large angular separations between the Sun and the area of the sky observed).\footnote{Interestingly, the other daysky spectra have $EW$s that are in good agreement with those measured in the asteroid spectra, which are point sources observed during night time. These daysky observations were made with the Tull spectrograph by letting scattered sunlight pass through a ``solar port'' which points towards the zenith, and a few hours before sunset, implying zenith-Sun angles between about 40 and 70 degrees. Although the solar spectral lines are expected to be distorted under these conditions, the impact on the $EW$ values is likely smaller. Note also that this effect is better seen at higher spectral resolution. The HET-HRS daysky spectrum has $R=120\,000$ whereas the Tull daysky spectra have $R=60\,000$.} Thus, before averaging the solar $EW$s, we corrected those of this spectrum by increasing them by 3\,\%, which brought them to close agreement (on the average) with the other solar spectra. We note also that this correction factor is in reasonably good agreement with that suggested by \cite{gray00}. After applying this correction, for each spectral line we averaged the $EW$s measured in all 10 spectra and adopted the mean value as the solar reference $EW$s, which are listed in column 5 of Table~\ref{t:linelist}.

Using our iron line list and a MARCS model atmosphere, we derive a solar iron abundance $\afe=7.46\pm0.06$ using only the \fei\ lines, and $\afe=7.44\pm0.08$ using only \feii\ lines. Our inferred solar iron abundance is thus in reasonably good agreement with estimates found elsewhere, in particular with that derived by \cite{asplund09:review} using a three-dimensional (3D) time-dependent hydrodynamic solar photosphere simulation. Note that 3D corrections to the 1D analysis (which we employ) are very small for iron lines in the solar case.

\subsubsection{Microturbulence}

As is common practice, the microturbulent velocity ($v_t$) was obtained by removing the correlation between iron abundance and reduced equivalent width ($\log(EW/\lambda)$) of \fei\ lines. For the solar reference we used the absolute abundances $A_\mathrm{\fei}$ and obtained $v_t=1.03$\,\kms. For the other stars we used the relative iron abundances, i.e., the $\feh$ values from the \fei\ line analysis. This procedure is more reliable than using absolute abundances because errors in the atomic data, and to a lesser extent in the model atmospheres, are minimized with our line-by-line differential analysis.

\begin{figure}
\includegraphics[bb=70 363 390 815,width=8.5cm]{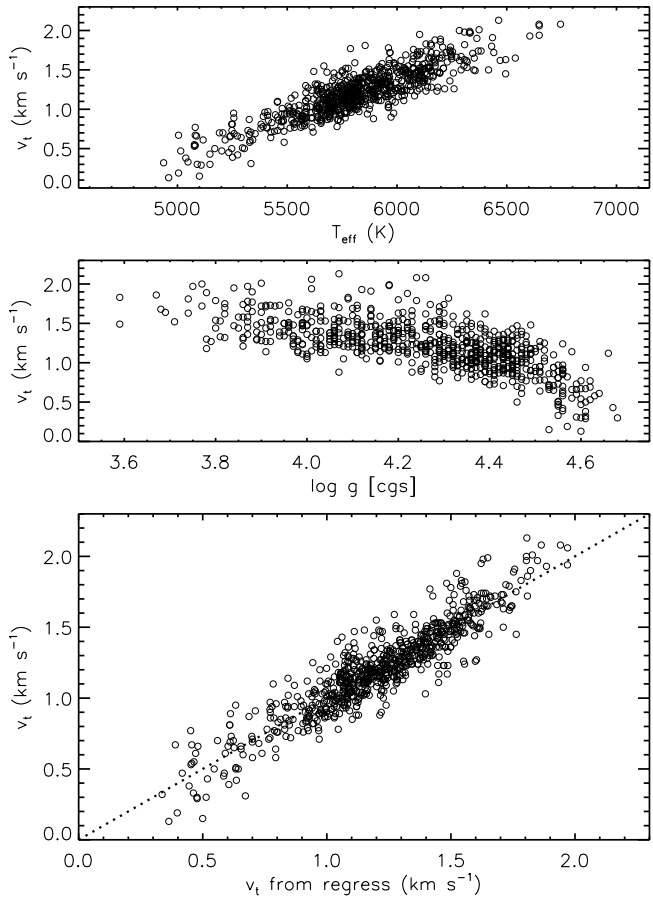}
\caption{Top panel: microturbulent velocity inferred from our analysis of \fei\ lines as a function of $\teff$. Middle panel: as in the top panel for $\logg$. Bottom panel: comparison of microturbulent velocities, inferred from the \fei\ lines, and obtained from the linear regression (Equation~\ref{eq:vt}). The dotted line shows the 1-to-1 correspondence. The stars excluded from the regression leading to Equation~\ref{eq:vt} are not shown in this figure.}
\label{f:vt}
\end{figure}

The resulting $v_t$ values correlate strongly with $\teff$ and $\logg$, as shown in Figure~\ref{f:vt} for the dwarf and subgiant stars. Giant stars (whose analysis is discussed later in Section~\ref{s:giants}) are excluded from Figure~\ref{f:vt} because they are at the cool $\teff$ end but have large $v_t$ values (from about 0.8 to 1.6\,\kms) compared to the cool K-dwarfs ($v_t<0.5$\,\kms). The $v_t$ values show only a weak dependence on $\feh$. The dependency of $v_t$ on stellar parameters that we obtain is consistent with the fact that the microturbulent velocity scales well with the strength of surface convection \cite[cf.][]{gray78,nordlund90}. We performed a linear regression for $v_t$ on these three parameters, using an iterative 2.5\,$\sigma$ clipping algorithm to remove outliers. We found the following mean relation:
\begin{eqnarray}
v_t\ (\mathrm{km\ s}^{-1}) & = & 1.163+7.808\times10^{-4}(\teff-5800\,\mathrm{K}) \nonumber \\
    &   & -0.494(\logg-4.30)-0.050\feh\,,
\label{eq:vt}
\end{eqnarray}
with a 1\,$\sigma$ error of 0.12\,\kms. The limits of applicability of this relation are: $\teff=4940-6750$\,K, $\logg=3.6-4.7$, and $\feh=-1.2$ to $\feh=+0.4$.

A comparison of $v_t$ values derived using Equation~\ref{eq:vt} with those obtained in the iron line analysis is shown in the bottom panel of Figure~\ref{f:vt}. For the stars that were removed from this regression due to the 2.5\,$\sigma$ clipping constrain (about 15\,\%\ of the sample), we adopted the $v_t$ values inferred from Equation~\ref{eq:vt} rather than that derived from their \fei\ line analysis. Note that $v_t$ is a parameter that could be severely affected by the few spectral lines on the extremes of the reduced equivalent width distribution and therefore the mean relation is more reliable for the outliers. The latter is also true for cool K-dwarfs ($\teff<5000$\,K), for which their $v_t$ values were inferred mainly by extrapolating Equation~\ref{eq:vt}. Note, however, that in none of these cases we obtained an unphysical (i.e., negative) $v_t$.

Compared to other $v_t$ parameterizations found in the literature, ours is in good qualitative agreement. For example, the formulas by \citet[][their Equation~9]{edvardsson93} and \citet[][their Equation~2]{allende04:s4n} both have a positive slope for the $\teff$ dependence, but a negative one for the $\logg$ dependence. Quantitatively, the $\teff$ dependence of the \cite{edvardsson93} formula for dwarf stars ($\logg=4.5$) is almost identical to ours (mean difference of only 0.03\,\kms\ with a $1\,\sigma$ standard deviation of 0.01\,\kms, with our $v_t$ values being smaller). The $\logg$ dependence, however, is stronger according to \cite{edvardsson93}, who predict $v_t\simeq1.9$\,\kms\ at $\logg=3.8$ while our parameterization suggests $v_t\simeq1.3$\,\kms. The formula by \cite{allende04:s4n}, on the other hand, implies even shallower $\teff$ and $\logg$ dependencies, but a $\logg$ dependence that is more similar to ours than that by \cite{edvardsson93}.

\subsubsection{Spectroscopic Equilibrium} \label{s:spectroscopic_equilibrium}

Ideally, the average iron abundance inferred from \fei\ and \feii\ lines separately should agree. In addition, the \fei\ abundances should not correlate with the excitation potential of the lines. In many studies, these conditions are used to fine-tune $\teff$, $\logg$, and $v_t$, given their sensitivity to those trends. We prefer to avoid that approach because it is strongly sensitive to model uncertainties whereas our atmospheric parameter determination uses techniques that are known to be less affected by those systematic errors. Fine-tuning by ionization/excitation balance, however, can be powerful for the analysis of stars which are very similar to the Sun, when the latter is used as reference star, because systematic errors are essentially cancelled out in a line-by-line differential analysis \citep{melendez09:twins,melendez12,ramirez09,ramirez11}.

\begin{figure}
\includegraphics[bb=65 382 390 785,width=8.5cm]{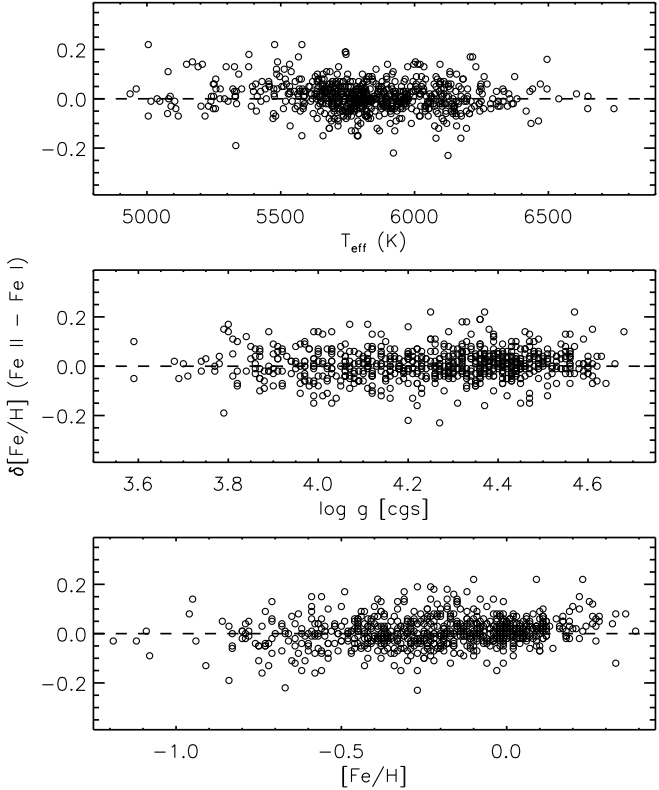}
\caption{Test of iron ionization equilibrium. The difference between the mean iron abundance determined from \feii\ and \fei\ lines is shown as a function of stellar parameters.}
\label{f:iron_ionization}
\end{figure}

The \feii\ minus \fei\ abundance difference from our iron abundance measurements is shown in Figure~\ref{f:iron_ionization}. This plot shows the stars from our ``main'' sample, but it does not include those for which $v_t$ was inferred from Eq.~\ref{eq:vt} (i.e., stars which were sigma-clipped from the fit are not plotted, which includes many with $\teff<5000$\,K). There is significant scatter in Figure~\ref{f:iron_ionization} but no obvious trends or offsets are observed. The average \feii--\fei\ difference is $0.01\pm0.06$\,dex. This is contrary to the case of R07, who found an overall offset of +0.06\,dex for the \feii\ abundances relative to \fei, as well as severe discrepancies for cool metal-rich dwarfs.

A more detailed inspection of the trends shown in Figure~\ref{f:iron_ionization}, particularly adding the stars with $v_t$ derived from Eq.~\ref{eq:vt}, which includes several cool K-dwarfs, reveals that there is a trend of higher \feii\ abundances for the cool metal-rich dwarfs in our data, as in R07. In fact, the mean \feii\ minus \fei\ difference for dwarf stars with $\teff<4800$\,K is $+0.09\pm0.07$, a number that increases to $+0.15\pm0.05$ if the sample selection is further restricted to stars with super-solar metallicity. This severe iron ionization problem is seen only in cool metal-rich K-dwarfs, and it has also been detected in other works, as will be discussed in detail in Section~\ref{s:fine-tuning}.

Excluding the stars mentioned above, we do not detect an overall offset in the iron abundances inferred from \fei\ and \feii\ lines separately at all $\teff$ values, as we found in our previous work. This is most likely due to the use of the improved IRFM $\teff$ scale by \cite{casagrande10}, who shifted the zero point by about +100\,K with respect to the $\teff$ scale by \cite{ramirez05b}. A +100\,K offset in the $\teff$ values used is almost the exact amount necessary to remove the \feii\ minus \fei\ abundance offset in the R07 data.

\begin{figure}
\includegraphics[bb=50 382 390 785,width=8.5cm]{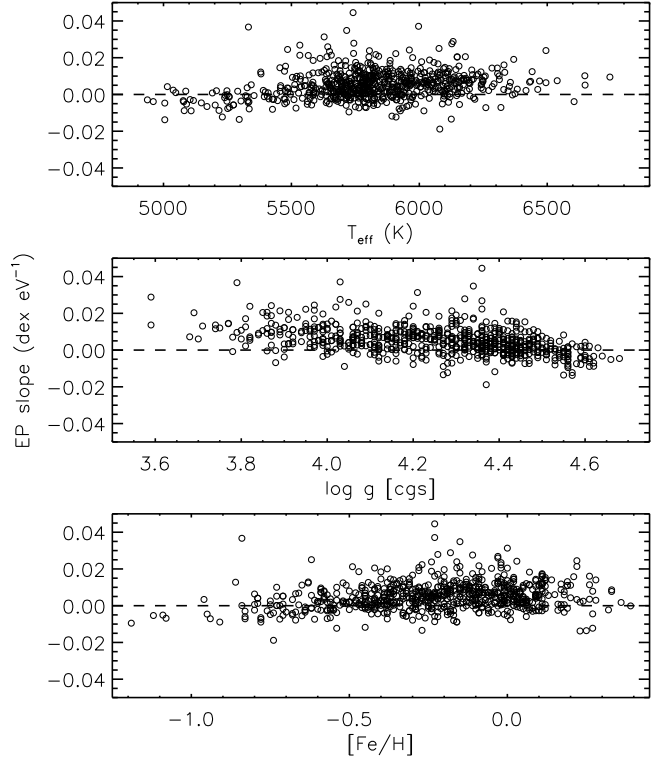}
\caption{Test of \fei\ excitation equilibrium. The slope of abundance versus excitation potential for \fei\ lines is shown as a function of stellar parameters.}
\label{f:ep_slope}
\end{figure}

Figure~\ref{f:ep_slope} shows the $\feh$ vs.\ EP slope for \fei\ lines. Our \fei\ linelist consists of features with EP from about 0 to 5 eV. The EP slope can be severely affected by one or two unreliable lines on the extremes of the EP distribution, which most likely explains the outliers in Figure~\ref{f:ep_slope}. The bulk of the data, however, reveals clear trends with all three parameters. Although excitation balance (i.e., no EP slope) appears to be satisfied around the solar values, the EP slope is clearly positive for relative low $\logg\lesssim4.3$ and negative for relative low $\feh\lesssim-1.0$. It is also low at cool temperatures ($\teff\lesssim5500$\,K). This implies that the $\teff$ values inferred using the condition of excitation balance are warmer than the effective temperatures obtained with the IRFM color calibrations for $\logg\lesssim4.3$ stars, but cooler for $\feh\lesssim-1.0$ and/or $\teff\lesssim5500$\,K stars.

Our atmospheric parameters do not satisfy perfectly the so-called spectroscopic equilibrium conditions, yet there has been a significant improvement with respect to R07 in that ionization balance is no longer an important issue, except at the cool metal-rich end. The discrepancy in the excitation equilibrium is likely due to systematic errors related to model atmospheres and departures from local thermodynamic equilibrium (LTE) in the spectral line formation calculations. Finding a solution to this problem is beyond the scope of this paper. We refer the reader to the recent works by \cite{bergemann12} and \cite{lind12} for a detailed investigation of these issues.

\subsection{Special Cases} \label{s:special}

\subsubsection{No Trigonometric Surface Gravity}

A number of our sample stars (11) are not included in the {\it Hipparcos} catalog and do not have ground-based measurements of their trigonometric parallax. They can be found in the Tycho catalog, but their parallaxes have large errors ranging from 30 to 100\,\%, with most of them showing an error in the parallax of at least 50\,\%. Thus, we cannot use the technique described in Section~\ref{s:logg} to derive their $\logg$ values with high precision. For this sub-sample of stars (the ``noplx'' sample) we obtained $\logg$ by forcing the mean \fei\ and \feii\ abundances to agree (ionization equilibrium). The effective temperatures were still obtained as described in Section~\ref{s:teff}.

We used the same approach for two stars (HIP\,5134 and HIP\,61272) which, even though they have an entry in the {\it Hipparcos} catalog, inspection of ionization equilibrium (\fei\ minus \feii\ abundance difference) and the wings of strong lines such as Mg\,\textsc{i}{\it b} suggest a severely erroneous parallax.

\subsubsection{Giants} \label{s:giants}

The \cite{casagrande10} color-$\teff$ relations apply only to dwarf and subgiant stars. No extension of this $\teff$ scale for giants has been published yet. Therefore, for a number of stars on the red giant branch (27, the ``giant'' sample) we determined $\teff$ using excitation equilibrium. However, we did not force the EP slope to be exactly zero but 0.01\,dex\,eV$^{-1}$ because this value is the average observed in the relatively low $\logg$ stars from the main sample (see Figure~\ref{f:ep_slope}). In this way, the $\teff$ values for giant stars are forced to be roughly on the same scale as that of the dwarf and subgiant stars. However, we cannot guarantee that this will be the case for all the giant stars, given that they cover a very wide range of radii, implying significantly different atmospheric structures compared to stars at the base of the red giant branch. Nevertheless, we note that for Arcturus (HIP\,69673), which is one of the stars with the lowest $\logg$ in our sample, we derive $\teff=4215\pm90$\,K, a value that is in good agreement, within 1\,$\sigma$, with that obtained in the very detailed analyses of this giant star by \citet[][$\teff=4290\pm30$\,K]{griffin99} and \citet[][$\teff=4286\pm30$\,K]{ramirez11:arcturus}.

\subsubsection{Other}

A few stars (18) presented several different difficulties when attempting to measure their atmospheric parameters with the techniques described above. In some cases very few lines were available, making the determination of $v_t$ extremely difficult. These could be not only very metal-poor stars, but also objects for which our spectra are of relatively low quality. In other cases, the published parallaxes returned unphysical $\logg$ values, as confirmed from inspection of the gravity-sensitive Mg\,\textsc{i}\,$b$ features at 5180\,\AA\ or from their odd location in the HR diagram. For these stars (the ``manual'' sample), whenever possible, we constrained $\teff$ and/or $\logg$ from excitation and ionization balance; otherwise we adopted the averages of values found in the literature, as compiled by \cite{soubiran10}.

\subsection{Catalog of Atmospheric Parameters}

\begin{deluxetable*}{lccrccrrl}
\tablewidth{0pc}
\tabletypesize{\scriptsize}
\tablecaption{Atmospheric Parameters\tablenotemark{1}}
\tablehead{\colhead{Star} & \colhead{$\teff$ (K)} & \colhead{$\logg$ [cgs]} & \colhead{$\feh$} & \colhead{$v_t$ (\kms)} & \colhead{cal.\tablenotemark{2}} & \colhead{$\feh_\mathrm{\fei}$} & \colhead{$\feh_\mathrm{\feii}$} & \colhead{sample}}
\startdata
HD 32071 & $5894\pm38$ & $4.23\pm0.12$ & $-0.34\pm0.04$ & 1.35 &  & $-0.34\pm0.04$ & $-0.33\pm0.07$ & noplx \\
HD 59490 & $5627\pm27$ & $4.49\pm0.11$ & $-0.17\pm0.04$ & 0.93 &  & $-0.17\pm0.03$ & $-0.17\pm0.08$ & noplx \\
HD 67163 & $5831\pm29$ & $4.26\pm0.09$ & $0.03\pm0.03$ & 1.26 &  & $0.03\pm0.03$ & $0.03\pm0.04$ & noplx \\
HD 82960 & $6450\pm129$ & $3.95\pm0.20$ & $0.04\pm0.04$ & 1.85 &  & $0.04\pm0.04$ & $0.02\pm0.07$ & manual \\
HD 130047 & $5590\pm33$ & $4.31\pm0.12$ & $-0.02\pm0.04$ & 0.91 &  & $-0.02\pm0.04$ & $-0.02\pm0.09$ & noplx \\
HD 144070 & $6532\pm87$ & $4.16\pm0.08$ & $0.02\pm0.10$ & 1.81 & + & $0.01\pm0.09$ & $0.07\pm0.13$ & main \\
HD 170058 & $5896\pm46$ & $4.21\pm0.14$ & $0.13\pm0.05$ & 1.29 &  & $0.13\pm0.04$ & $0.13\pm0.10$ & noplx \\
HD 171029 & $5562\pm54$ & $3.85\pm0.21$ & $-0.36\pm0.07$ & 0.97 &  & $-0.36\pm0.07$ & $-0.36\pm0.08$ & noplx \\
HD 183490 & $5759\pm43$ & $3.81\pm0.14$ & $0.06\pm0.05$ & 1.25 &  & $0.06\pm0.05$ & $0.07\pm0.05$ & noplx \\
HD 213746 & $5813\pm49$ & $4.12\pm0.18$ & $0.05\pm0.06$ & 1.34 &  & $0.05\pm0.05$ & $0.05\pm0.11$ & noplx \\
HD 223723 & $5946\pm41$ & $4.28\pm0.10$ & $0.01\pm0.03$ & 1.28 &  & $0.01\pm0.04$ & $-0.01\pm0.02$ & noplx \\
HIP 171 & $5510\pm66$ & $4.46\pm0.01$ & $-0.76\pm0.06$ & 0.96 &  & $-0.76\pm0.05$ & $-0.80\pm0.11$ & main \\
HIP 348 & $5746\pm55$ & $4.38\pm0.06$ & $-0.19\pm0.04$ & 1.13 &  & $-0.20\pm0.03$ & $-0.18\pm0.07$ & main \\
HIP 394 & $5635\pm37$ & $3.78\pm0.07$ & $-0.48\pm0.04$ & 1.24 &  & $-0.48\pm0.03$ & $-0.47\pm0.06$ & main \\
HIP 475 & $5836\pm72$ & $4.35\pm0.05$ & $-0.34\pm0.06$ & 1.18 &  & $-0.33\pm0.05$ & $-0.41\pm0.10$ & main \\
HIP 493 & $5960\pm44$ & $4.41\pm0.03$ & $-0.20\pm0.04$ & 1.16 &  & $-0.20\pm0.03$ & $-0.20\pm0.07$ & main \\
HIP 522 & $6251\pm44$ & $4.21\pm0.02$ & $0.05\pm0.05$ & 1.56 & + & $0.04\pm0.05$ & $0.07\pm0.06$ & main \\
HIP 530 & $5866\pm40$ & $3.90\pm0.05$ & $-0.01\pm0.05$ & 1.36 &  & $-0.02\pm0.03$ & $0.03\pm0.09$ & main \\
HIP 544 & $5458\pm40$ & $4.52\pm0.02$ & $0.14\pm0.06$ & 0.78 & + & $0.13\pm0.04$ & $0.17\pm0.13$ & main \\
HIP 656 & $5805\pm39$ & $3.82\pm0.06$ & $-0.24\pm0.06$ & 1.32 &  & $-0.25\pm0.05$ & $-0.15\pm0.07$ & main \\
\ldots & \ldots & \ldots & \ldots & \ldots & \ldots & \ldots & \ldots & \ldots 
\enddata
\tablenotetext{1}{Table~\ref{t:pars} is published in its entirety in the electronic edition of the Astrophysical Journal. A portion is shown here for guidance regarding its form and content.}
\tablenotetext{2}{A + sign is written in this column if the adopted microturbulent velocity for the star was obtained from the linear regression (Equation~\ref{eq:vt}).}
\label{t:pars}
\end{deluxetable*}

Our final atmospheric parameters are listed in Table~\ref{t:pars}. Results from multiple spectra of the same star have already been averaged in this table. The latter was applied to 68 objects, with an average spectrum-to-spectrum $1\,\sigma$ scatter of less than 0.02\,dex in $\feh$. For $\logg$ and $\teff$, this scatter is even smaller: 0.01\,dex and 2\,K, respectively. This is due to the fact that $\logg$ and $\teff$ are derived almost independently from the spectrum; $\feh$ is an input quantity that has only a minor impact on their derivation. Photometric and astrometric errors dominate the uncertainties in $\teff$ and $\logg$, respectively. Since the spectrum-to-spectrum scatter of the derived stellar parameters is small compared to the internal error (see below), we conclude that the heterogeneity of our spectroscopic data set has a negligible impact on our results.

The median errors of the stellar parameters (internal only, i.e., not including estimates of systematic errors) are: $\Delta\teff=44$\,K, $\Delta\logg=0.04$, and $\Delta\feh=0.05$ (the way in which the individual errors were determined are described in previous sections).

\begin{deluxetable*}{lcccccccccc}
\tablewidth{0pc}
\tabletypesize{\scriptsize}
\tablecaption{Age and Mass\tablenotemark{1}}
\tablehead{\colhead{Star} & \colhead{Age (Gyr)} & \colhead{$-1\sigma$} & \colhead{$+1\sigma$} & \colhead{$-2\sigma$} & \colhead{$+2\sigma$} & \colhead{Mass ($M_\odot$)} & \colhead{$-1\sigma$} & \colhead{$+1\sigma$} & \colhead{$-2\sigma$} & \colhead{$+2\sigma$}}
\startdata
  HD 32071& 9.09& 7.12& 9.68& 3.92&10.75  & 0.942& 0.929& 0.992& 0.906& 1.054\\
  HD 59490& 6.43& 3.44&10.64& 1.15&12.89  & 0.918& 0.895& 0.936& 0.877& 0.956\\
  HD 67163& 7.31& 5.64& 7.74& 3.31& 8.46  & 1.032& 1.016& 1.062& 1.000& 1.096\\
  HD 82960& 2.01& 1.71& 2.89& 1.06& 3.84  & 1.313& 1.313& 1.692& 1.215& 1.873\\
 HD 130047&11.65& 7.10&12.17& 2.88&13.39  & 0.936& 0.914& 0.957& 0.898& 0.993\\
 HD 144070& 2.14& 1.73& 2.55& 1.12& 3.00  & 1.384& 1.334& 1.453& 1.283& 1.520\\
 HD 170058& 5.22& 4.01& 6.29& 1.81& 7.20  & 1.102& 1.078& 1.195& 1.037& 1.281\\
 HD 171029& 6.89& 4.37&12.53& 2.24&14.36  & 0.940& 0.906& 1.242& 0.851& 1.535\\
 HD 183490& 3.90& 3.24& 6.60& 2.21& 8.36  & 1.242& 1.139& 1.446& 1.052& 1.637\\
 HD 213746& 7.16& 5.04& 8.03& 2.65& 9.20  & 1.058& 1.013& 1.158& 0.982& 1.323\\
 HD 223723& 5.70& 3.96& 6.22& 1.83& 7.04  & 1.078& 1.053& 1.119& 1.031& 1.169\\
   HIP 171&14.46&12.65&14.56&10.92&14.78  & 0.759& 0.751& 0.771& 0.744& 0.787\\
   HIP 348& 8.67& 5.28&10.20& 2.63&12.19  & 0.930& 0.913& 0.963& 0.891& 0.993\\
   HIP 394& 5.00& 3.93& 7.22& 3.01& 9.34  & 1.132& 1.053& 1.257& 0.977& 1.369\\
   HIP 475& 8.75& 7.09&11.47& 4.45&13.32  & 0.923& 0.891& 0.954& 0.863& 0.989\\
   HIP 493& 5.39& 3.21& 6.06& 1.78& 7.21  & 1.004& 0.986& 1.034& 0.966& 1.057\\
   HIP 522& 3.00& 2.70& 3.40& 2.37& 3.63  & 1.244& 1.232& 1.258& 1.222& 1.274\\
   HIP 530& 4.90& 4.40& 5.60& 3.85& 6.16  & 1.222& 1.184& 1.284& 1.142& 1.333\\
   HIP 544& 3.51& 1.59& 6.23& 0.46& 8.80  & 0.950& 0.919& 0.976& 0.892& 0.995\\
   HIP 656& 4.19& 3.51& 5.39& 2.90& 6.51  & 1.254& 1.178& 1.346& 1.109& 1.429\\
\ldots & \ldots & \ldots & \ldots & \ldots & \ldots & \ldots & \ldots & \ldots & \ldots & \ldots 
\enddata
\tablenotetext{1}{Table~\ref{t:age} is published in its entirety in the electronic edition of the Astrophysical Journal. A portion is shown here for guidance regarding its form and content.}
\label{t:age}
\end{deluxetable*}

In Table~\ref{t:pars}, a flag for $v_t$ tells whether the microturbulence was derived from the \fei\ analysis or from the linear regression (Equation~\ref{eq:vt}). Similarly, we include a column with information on how exactly were the atmospheric parameters derived (column ``sample''). Our adopted stellar masses and ages, derived as described in Section~\ref{s:logg}, are given in Table~\ref{t:age}. For stars with no trigonometric parallax available, hence with unknown absolute magnitude, the exact same mathematical procedure explained in Section~\ref{s:logg} was used, replacing $M_V$, the observed absolute magnitude, with the spectroscopically measured $\logg$, and $m$, the absolute magnitude from the models, with the $\logg$ values from the isochrones.

\subsection{Oxygen Abundance}

As shown before, the \oi\ triplet at 777\,nm is a very strong feature in solar-type stars, easily detectable in high resolution, high signal-to-noise spectra. The equivalent widths of these three lines can be measured with high precision for most nearby FGK-type stars (cf.\ Figure~\ref{f:triplet_sample}). Accurate transition probabilities, based on quantum mechanical calculations, are available from \cite{hibbert91}, \cite{butler91}, and \cite{biemont92}. These published values are reasonably consistent with each other and we therefore adopted their averages: $\log gf=0.352,0.223,0.002$ for $\lambda=7771.9,7774.2,7775.4$\,\AA, respectively.

However, it has long been known that the oxygen infrared triplet lines form in conditions that are not well reproduced under the LTE assumption \cite[e.g.,][]{eriksson79,kiselman93}. Statistical equilibrium calculations show that in the solar case, the LTE oxygen abundance inferred from the \oi\ 777\,nm triplet is too high by $\sim0.2$\,dex ($\sim50$\,\%) relative to other oxygen features less affected by model uncertainties \cite[e.g.,][]{kiselman93,takeda94,ramirez07,fabbian09}. Non-LTE effects are stronger for warmer and lower surface gravity stars, owing to the stronger radiation fields and lower densities, which make radiative transitions more important relative to collisional excitation. Nevertheless, even for the coolest dwarf stars in our sample, non-LTE effects can introduce (predicted) errors of almost 0.1\,dex. On the other hand, the LTE oxygen abundances of the warmest stars in our sample are overestimated by about 0.5\,dex, i.e., by more than a factor of 2.

In R07, we used an oxygen model atom composed of 54 levels plus the continuum and 242 transitions, constructed by \cite{allende03a}, along with the computer codes TLUSTY and SYNSPEC \citep{hubeny88,hubeny95} to calculate level populations for the oxygen atom and emergent flux synthetic \oi\ 777\,nm triplet lines in a grid of stellar atmospheres. Non-LTE corrections to the abundances derived using the LTE approximation were then computed for this grid using a wide range of oxygen abundances. We used this grid of non-LTE corrections to interpolate linearly to the stellar parameters of our sample. The grid has a fine spacing in stellar parameters and interpolation errors are modest. R07 estimate that interpolation within the grid is consistent with non-LTE corrections computed for each star individually within 0.02\,dex, but in most cases the agreement is better than within 0.01\,dex.

One important limitation of the non-LTE calculations employed in this work is the neglect of inelastic collisions with neutral hydrogen in the statistical equilibrium calculations. They tend to bring the predicted line strengths closer to their LTE values, thus reducing the size of the non-LTE corrections. Collisions with neutral hydrogen can be taken into account in the non-LTE calculations using the modified classical Thomson formula by \cite{drawin68} for the cross-sections. As pointed out by \cite{steenbock84} and \cite{lambert93}, among others, Drawin's formula likely provides only an order of magnitude estimate, and it is therefore common to scale its prediction with an empirical $S_\mathrm{H}$ factor. It is expected that this number depends on the type of transition \citep{lambert93,barklem11}, but it is often assumed constant.

Center-to-limb variation (CLV) observations of the triplet can be used to constrain $S_\mathrm{H}$, an exercise that is obviously limited to the Sun \cite[e.g.,][]{allende04:clv}. Using a three-dimensional hydrodynamic simulation of the solar photosphere, \cite{pereira09a} find the best agreement between model and observation for the CLV of the triplet if an $S_\mathrm{H}\simeq1$ is adopted. \cite{ramirez06}, on the other hand, find that the three lines of the triplet in the spectrum of the standard SDSS star BD\,+17\,4708 ($\feh\simeq-1.7$) are most consistently fitted using $S_\mathrm{H}\simeq10$, although that work uses one-dimensional static models and the oxygen abundance inferred for this well-known moderately metal-poor star with that choice of $S_\mathrm{H}$ appears higher than usual values for halo stars.

Further investigation on the impact of inelastic collisions with neutral hydrogen on non-LTE calculations is clearly needed to better constrain $S_\mathrm{H}$ as well as its possibly important variation across stellar parameter space. Note, however, that the detailed quantum mechanical calculations by \cite{barklem11} show that the physics of excitation by H collisions is not well reproduced by Drawin's formula, which could result in poorly predicted {\it relative} rates for collisional transitions.

In this work, we did not take into account collisions with neutral H. Therefore, there is no $S_\mathrm{H}$ factor associated with our non-LTE calculations. Nevertheless, as explained in R07, by analyzing the non-LTE oxygen abundances inferred for each of the triplet lines, we found that in order to obtain consistent abundances from the three lines (on the average for all stars) small corrections were needed. The non-LTE oxygen abundances from the 7771.9\,\AA\ line need to be reduced by 0.036\,dex while those from the 7775.4\,\AA\ need to be reduced by 0.018\,dex, keeping the abundance from the weakest line constant. The latter is expected to be less affected by collisions with neutral H, while the direction of these empirical corrections is consistent with them being due to the neglect of this effect, which should be more important for the strongest line which forms in higher atmospheric layers with lower H densities. Thus, the impact of collisions with neutral H is reduced, although certainly not fully removed, with these empirical corrections, which we also applied in this work. Interestingly, the more recent, accurate non-LTE calculations by \cite{fabbian09}, which include collisions with neutral H adopting $S_\mathrm{H}=1$, result in a non-LTE abundance correction for the Sun that is very similar to that from our work.

\begin{figure}
\includegraphics[bb=60 360 390 750,width=8.5cm]{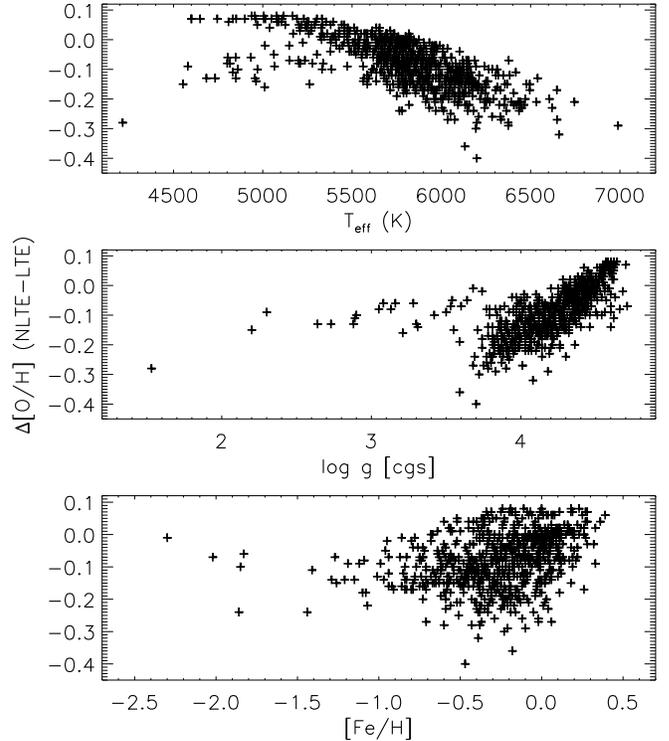}
\caption{Non-LTE corrections applied to the LTE [O/H] values inferred from our data as a function of stellar atmospheric parameters.}
\label{f:nlte}
\end{figure}

We derived LTE oxygen abundances using the \verb"abfind" driver in MOOG. Given the strong sensitivity of the non-LTE effects described above to the stellar atmospheric parameters, and the wide range of $\teff,\logg,\feh$ values covered by our sample, we could not rely only on a solar differential analysis, as it is sometimes done. Figure~\ref{f:nlte} shows the difference in the [O/H] values derived using non-LTE and LTE oxygen abundances. Although the differences are close to zero around the solar parameters, even within a relatively small range of 100\,K in $\teff$ and 0.1\,dex in $\logg$, the $\Delta\mathrm{[O/H]}$ differences vary between about $-0.1$ and $+0.1$\,dex. These corrections are sizable: non-LTE effects must be properly taking into account before using oxygen abundances inferred from the 777\,nm triplet to interpret Galactic chemical evolution patterns.

\begin{deluxetable*}{lccccccrr}
\tablewidth{0pc}
\tabletypesize{\scriptsize}
\tablecaption{Oxygen Abundance\tablenotemark{1}}
\tablehead{\colhead{Star} & \colhead{} & \colhead{$A_\mathrm{O,LTE}$} & \colhead{} & \colhead{} & \colhead{$A_\mathrm{O,NLTE}$} & \colhead{} & \colhead{$\mathrm{[O/H]}_\mathrm{LTE}$} & \colhead{$\mathrm{[O/H]}_\mathrm{NLTE}$} \\ \colhead{} & \colhead{7772\,\AA} & \colhead{7774\,\AA} & \colhead{7775\,\AA} & \colhead{7772\,\AA} & \colhead{7774\,\AA} & \colhead{7775\,\AA} & \colhead{} & \colhead{}}
\startdata
  HD 32071&8.71&8.69&8.71&8.47&8.45&8.50& $-0.08\pm0.01$& $-0.18\pm0.02$\\
  HD 59490&8.67&8.66&8.71&8.54&8.55&8.59& $-0.10\pm0.03$& $-0.09\pm0.02$\\
  HD 67163&8.88&8.87&8.89&8.70&8.69&8.73& $ 0.10\pm0.02$& $ 0.05\pm0.02$\\
  HD 82960&9.12&9.09&9.09&8.75&8.73&8.76& $ 0.32\pm0.01$& $ 0.09\pm0.01$\\
 HD 130047&8.96&8.94&8.94&8.80&8.80&8.81& $ 0.16\pm0.01$& $ 0.15\pm0.01$\\
 HD 144070&9.31&9.19&9.12&8.94&8.85&8.81& $ 0.42\pm0.07$& $ 0.21\pm0.05$\\
 HD 170058&9.03&9.04&9.05&8.83&8.83&8.87& $ 0.26\pm0.02$& $ 0.19\pm0.02$\\
 HD 171029&8.90&8.90&8.96&8.63&8.62&8.70& $ 0.14\pm0.04$& $-0.00\pm0.04$\\
 HD 183490&9.10&9.09&9.03&8.81&8.81&8.80& $ 0.29\pm0.02$& $ 0.15\pm0.01$\\
 HD 213746&8.84&8.87&8.88&8.64&8.67&8.71& $ 0.08\pm0.03$& $ 0.02\pm0.03$\\
 HD 223723&8.94&8.90&8.94&8.72&8.70&8.75& $ 0.14\pm0.02$& $ 0.07\pm0.02$\\
   HIP 171&8.54&8.44&8.52&8.36&8.27&8.36& $-0.26\pm0.02$& $-0.30\pm0.02$\\
   HIP 348&8.73&8.72&8.68&8.57&8.56&8.54& $-0.07\pm0.01$& $-0.10\pm0.01$\\
   HIP 394&8.71&8.64&8.64&8.42&8.37&8.40& $-0.11\pm0.02$& $-0.25\pm0.01$\\
   HIP 475&8.55&8.53&8.53&8.36&8.36&8.38& $-0.25\pm0.01$& $-0.30\pm0.01$\\
   HIP 493&8.69&8.67&8.58&8.50&8.49&8.43& $-0.14\pm0.04$& $-0.18\pm0.03$\\
   HIP 522&9.11&9.02&8.99&8.81&8.75&8.76& $ 0.26\pm0.04$& $ 0.12\pm0.02$\\
   HIP 530&9.10&9.06&9.03&8.80&8.78&8.77& $ 0.28\pm0.02$& $ 0.13\pm0.01$\\
   HIP 544&8.98&8.93&8.89&8.88&8.83&8.80& $ 0.15\pm0.02$& $ 0.18\pm0.03$\\
   HIP 656&8.98&8.93&8.87&8.64&8.62&8.59& $ 0.14\pm0.03$& $-0.04\pm0.02$\\
\ldots & \ldots & \ldots & \ldots & \ldots & \ldots & \ldots & \ldots & \ldots
\enddata
\tablenotetext{1}{Table~\ref{t:oxygen} is published in its entirety in the electronic edition of the Astrophysical Journal. A portion is shown here for guidance regarding its form and content.}
\label{t:oxygen}
\end{deluxetable*}

Our LTE and non-LTE oxygen abundances, both absolute and differential with respect to the solar oxygen abundance, as inferred from the \oi\ 777\,nm triplet lines, are listed in Table~\ref{t:oxygen}. The [O/H] errors given there correspond to only the line-to-line scatter for the three lines of the 777\,nm triplet. The error in [O/H] due to uncertainties in stellar parameters is dominated by $\Delta\teff$, which has a median value of 44\,K. Propagating this error into our calculation of [O/H] for a Sun-like star results in an error of 0.035\,dex. Similarly, the propagation of the median error in $\logg$ and $\feh$ into the [O/H] determination of a Sun-like star results in errors of only 0.001 and 0.002\,dex, respectively. Conservatively, hereafter we adopt a median [O/H] error of $\Delta\mathrm{[O/H]}=0.04$\,dex due to uncertainties in the stellar parameters.

Since the median line-to-line scatter for the non-LTE corrected [O/H] abundances is 0.02\,dex, we conclude that our [O/H] values have a median error of about 0.045\,dex. Estimating the error in [O/Fe] is not straightforward, because both the oxygen and iron abundances depend on the stellar parameters. Adding the oxygen and iron abundance errors in quadrature we obtain $\Delta\ofe\simeq0.07$\,dex, but this is an overestimated error. Based on a number of test calculations made taking into account the changes in the derived abundances due to artificial shifts in stellar parameters which correspond to their median errors, we find that for most of our sample stars, $\Delta\ofe\simeq0.05-0.06$\,dex. Note that we adopt $\feh$ values inferred from the \feii\ line analysis. For a Sun-like star, a change of $\pm44$\,K in $\teff$ implies a change in $\feh$ of $\pm0.03$\,dex if using \fei\ lines, but $\mp0.015$\,dex if using \feii. The latter goes in the same direction with respect to the change in [O/H], implying that the $\teff$ error compensates somewhat in Fe and O separately to make the $\ofe$ uncertainty smaller.

For the Sun, we derive an oxygen abundance $A_\mathrm{O}=8.77$, with a line-to-line scatter of only 0.011\,dex. After applying the non-LTE corrections, this value reduces to $A_\mathrm{O}=8.64$. Our non-LTE solar oxygen abundance is in good agreement (within 0.05\,dex) with that measured by \cite{asplund09:review} using a variety of spectroscopic indicators in addition to the 777\,nm \oi\ triplet, including a number of [\oi] lines which are expected to be formed in LTE. Moreover, our solar oxygen abundance is within 0.01\,dex of the abundance inferred from the triplet lines by \cite{asplund04}.

\section{OXYGEN ABUNDANCE PATTERNS} \label{s:trends}

\subsection{Fine-tuning the Sample} \label{s:fine-tuning}

Even though significant improvements to the determination of stellar parameters and oxygen abundances have been made since our previous publication on this topic (R07), inspection of abundance trends, in particular the [O/Fe] versus [Fe/H] relation, revealed a number of outliers and small subsamples of stars that added significant scatter to the mean relations observed. This suggests that there are systematic errors beyond our control that are still affecting our results, albeit only for a limited number of stars.

\begin{figure}
\includegraphics[bb=70 370 390 1055,width=8.9cm]{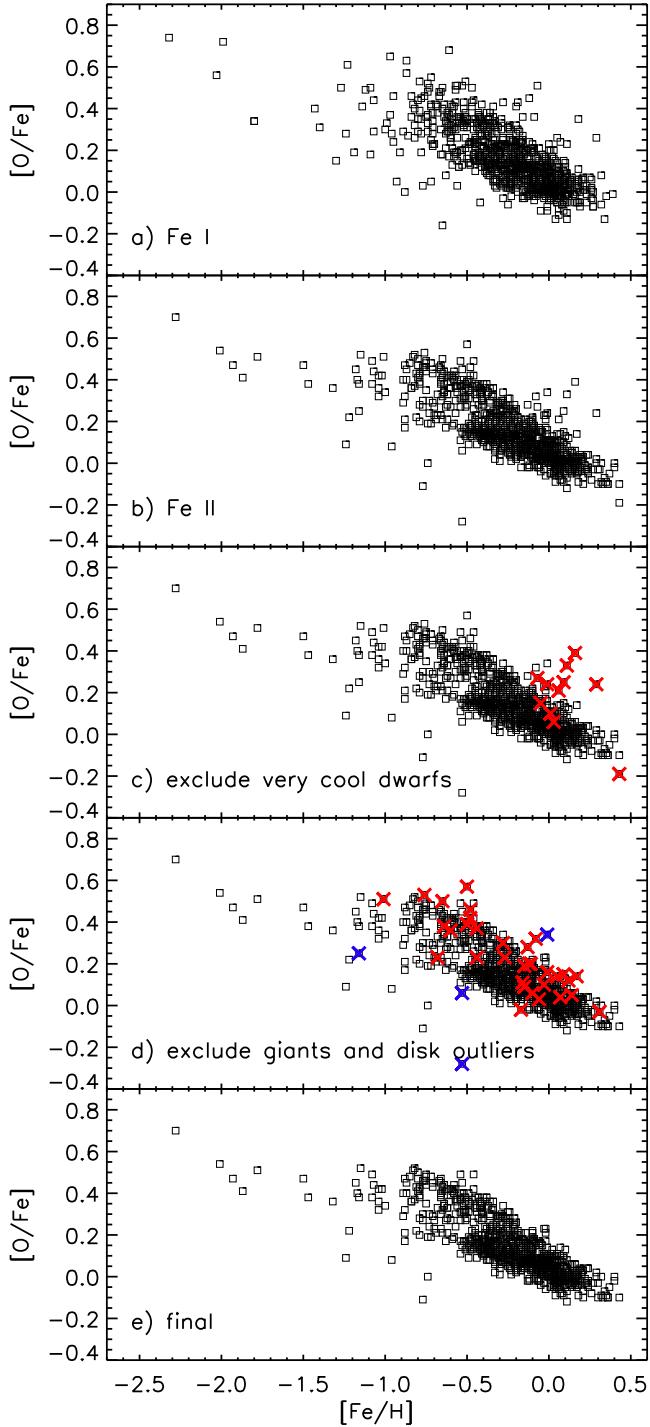}
\caption{{\bf a)} [O/Fe] versus [Fe/H] for all stars in our sample; [Fe/H] is based on \fei\ lines only. {\bf b)} As in a) for [Fe/H] based on \feii\ lines. {\bf c)} As in b); very cool dwarf stars are marked with large crosses and excluded from further analysis. {\bf d)} As in c); giant stars and objects which clearly depart from the mean thin/thick disk trends are marked with larger crosses and excluded from further analysis. {\bf e)} Final [O/Fe] versus [Fe/H] data adopted in this work.}
\label{f:ofe_all}
\end{figure}

In Figure~\ref{f:ofe_all}a we show the [O/Fe] versus [Fe/H] relation derived using the [Fe/H] values inferred from the \fei\ lines only. In Figure~\ref{f:ofe_all}b we show the same relation, but for \feii. Note that Figure~\ref{f:ofe_all} does not discriminate our sample stars according to their kinematics. In both Figures~\ref{f:ofe_all}a and b the well-known general trend of the Galactic disk is observed \cite[e.g.,][]{edvardsson93}, i.e., increasing [O/Fe] abundance ratios with decreasing [Fe/H], reaching a nearly constant value of $\ofe\simeq0.45$ at $\feh\lesssim-1$, where it connects with the halo. However, the star-to-star scatter is smaller when using [Fe/H] values from the \feii\ lines alone. In the majority of our sample stars, \feii\ is the dominant species of iron, which makes the \feii\ lines less sensitive to uncertainties in the stellar parameters and possibly also to modeling errors. Thus, although based on a fewer number of features, therefore implying larger random errors, the [Fe/H] values from the \feii\ analysis are more robust and they are adopted hereafter as the stellar metallicities.

Severe problems have been reported for both the iron and \oi\ 777\,nm triplet oxygen abundance determinations in cool K-type dwarf stars, particularly at high metallicities. Chemical abundance measurements of stars in young open clusters show that for $\teff\lesssim5000$\,K, the iron and oxygen abundances diverge towards very high values as $\teff$ decreases \cite[e.g.,][]{yong04,morel04,schuler06:hyades}. However, the precise way in which these abundances diverge from the expected values seems to depend on properties of the cluster such as metallicity and age. \cite{ramirez08:thesis} showed that this discrepancy is unlikely to be related to uncertainties in the modeling of stellar atmospheres due to the simplicity in the treatment of surface convection. The impact of chromospheric activity, starspots, and non-LTE effects remains to be fully explored and it could help understanding these problems. In Figure~\ref{f:ofe_all}c, stars with $\teff<5100$\,K, $\logg>4.4$, and $\feh>-0.1$ are highlighted with large crosses (11 stars). With few exceptions, these objects are well above the mean trend. Given their very uncertain atmospheric parameters and iron/oxygen abundances, we exclude these stars from further analysis.

As noted before, our sample consists mainly of dwarf and subgiant stars. Nevertheless, a non-negligible number of giant stars are also included (35, defined here by: $\teff<5500$\,K and $\logg<4.0$). They are highlighted with large crosses in Figure~\ref{f:ofe_all}d (red in the color version). As will be shown later, Galactic disk stars seem to follow distinct abundance patterns depending on their kinematics. Detailed inspection of the giant star data in Figure~\ref{f:ofe_all}d, separating the stars into thin- and thick-disk members as described in Section~\ref{s:kinematics}, shows a pattern nearly identical to that of the dwarf and subgiant stars, but shifted slightly in both [O/Fe] and [Fe/H]. Thus, adding the giant stars to the dwarf/subgiant data increases the star-to-star scatter in the oxygen abundance trends. One could attempt to put the giant star data into the dwarf/subgiant scale by applying constant offsets in [O/Fe] and [Fe/H], but this type of empirical procedure is risky; systematic errors do not always exhibit such linear behavior. Indeed, \cite{alves-brito10} have demonstrated that the practice of combining dwarf/subgiant elemental abundance data with giant star data can lead to spurious results. Given that our methods are optimized for the study of dwarf and subgiant stars, hereafter we exclude from our work objects on the red giant branch. We should note, however, that for the particular case of oxygen, the dwarf/giant discrepancies may be present when using the 777\,nm \oi\ triplet lines and not other features due to oxygen. For example, in their extensive abundance analysis of stars in the local region, \cite{luck06,luck07} find that their oxygen abundance trends for dwarf and giant stars are equivalent. They used only the [\oi] line at 630.0\,nm for their oxygen abundance analysis. This feature is expected to be less model dependent than the infrared triplet, but its analysis is complicated by the fact that it is a weak feature blended by a Ni line.

Even after eliminating stars with uncertain parameters and abundances, or inconsistent abundance trends relative to the bulk of our data, a few disk objects (four) were found as outliers relative to the main disk [O/Fe] versus [Fe/H] trend: HIP\,4039, HIP\,22060, HIP\,28103, and HIP\,49988. These objects are marked with large crosses in Figure~\ref{f:ofe_all}d (blue in the color version). It is possible that their abundances are peculiar, but we cannot rule out the possibility that a cool faint companion is affecting their photometry while not appearing obvious in the composite visible spectra. Very uncertain parallaxes could also be partially responsible for these outliers. Hereafter, these objects are also excluded from our work. The sample finally adopted for further analysis (consisting of 775 stars, or 94\,\% of all our sample stars) is shown in Figure~\ref{f:ofe_all}e. The overall pattern of increasing [O/Fe] with decreasing [Fe/H] for disk stars now appears very clean. Its detailed structure is discussed below in Sections~\ref{s:disk_trends} to \ref{s:disk_duality}. The few objects with low $\ofe\sim0.0$ at $\feh\sim-1.0$ are halo stars whose somewhat peculiar abundances are discussed in Section~\ref{s:halo}.

\subsection{Oxygen Abundance of Solar-metallicity Stars}

As mentioned before, one of the most important improvements made with respect to our previous work (R07) is the use of the updated IRFM $\teff$ scale by \cite{casagrande10}. The average +100\,K offset in $\teff$ between that $\teff$ scale and that by \cite{ramirez05a,ramirez05b}, which was used in R07, has resulted in important offsets in [O/H] and [Fe/H]. This is particularly interesting when looking at the [O/Fe] of stars with near solar metallicity ($\feh=0$). In R07, these stars seemed to have an average $\ofe\simeq0.1$ (see, for example, their Figure~8), leaving the Sun as a somewhat peculiar star regarding its oxygen content. Although not completely beyond the thin-disk [O/Fe] dispersion at that $\feh$, the Sun appeared to have a marginally low oxygen-to-iron abundance ratio relative to other stars with similar parameters. Using our newly derived iron and oxygen abundances, however, the location of the Sun is in much better agreement with that of most disk stars, as suggested by Figure~\ref{f:ofe_all}e.

The works by \cite{melendez09:twins} and \cite{ramirez09} on extremely high precision elemental abundances of solar twin stars, objects which have stellar parameters so similar to the Sun that systematic errors in the abundance analysis are minimized, show that the Sun in fact has a slightly higher oxygen-to-iron abundance ratio relative to its twins \cite[see also][]{ramirez10}. With abundance errors of about 0.01\,dex (quantified as the standard error for the mean abundance ratios of small samples of solar twins), they find that the mean [O/Fe] of solar twins is slightly sub-solar. \cite{melendez09:twins} derive an average $\ofe=-0.033\pm0.011$ for 11 solar twins while \cite{ramirez09} find $\ofe=-0.015\pm0.006$ using 22 of those objects. Restricting our sample to objects with $\teff=5777\pm100$\,K, $\logg=4.44\pm0.1$, and $\feh=0.0\pm0.1$, which is the practical definition of solar twin by \cite{ramirez10}, we find 47 stars with a mean $\ofe=0.02\pm0.04$, which is consistent with zero within our uncertainties, but also with the slightly low solar oxygen abundance suggested by \cite{melendez09:twins} and \cite{ramirez09}. Moreover, in their photometric studies of solar twin stars, \cite{melendez10}, \cite{ramirez12_suncolor}, and \cite{casagrande12} have concluded that the \cite{casagrande10} $\teff$ scale should be corrected by about +20\,K in order to be in perfect agreement with the highly-precise solar colors derived using model-independent methods. An increase of +20\,K in $\teff$ implies a decrease of about 0.01\,dex in [O/Fe], bringing our average solar twin [O/Fe] abundance ratio even closer to the values found by \cite{melendez09:twins} and \cite{ramirez09}.

\subsection{Disk: Kinematic Abundance Trends} \label{s:disk_trends}

As demonstrated by a number of works, oxygen abundance patterns are different for nearby disk stars with dissimilar kinematic properties. Galactic disk stars with cold (warm) kinematics are associated with the thin (thick) disk. It is generally accepted that the thin-disk stars are, as a sample, more iron-rich than thick-disk stars, and have lower [O/Fe] ratios relative to thick-disk stars of similar [Fe/H] \cite[e.g.,][]{gratton96,gratton00,prochaska00:thick-disk,tautvaivsiene01,bensby04,zhang06:thick-disk,ramirez07}. While the latter seems to be the case below $\feh\simeq-0.2$, for higher $\feh$ values, thin- and thick-disk star [O/Fe] versus [Fe/H] trends appear to converge to a common relation.

The difference in [O/Fe] abundance ratios of thin- and thick-disk stars described above has been attributed to their separate origin. Thick-disk stars are thought to have been born from material enriched mainly by Type\,II supernovae (SNII) yields, which are high in O/Fe \cite[e.g.,][]{woosley95}. Thin-disk stars, on the other hand, may have originated from gas contaminated also by Type\,Ia supernovae (SNIa) yields, which have high Fe/O \cite[e.g.,][]{iwamoto99}. The latter would increase the $\feh$ values while quickly decreasing the [O/Fe] abundance ratios. Thus, the knee connecting the thick disk [O/Fe]--[Fe/H] trend with that of the thin disk at high $\feh$ could also be a signature of SNIa pollution, as argued by \cite{bensby04}. Note, however, that the slightly decreasing [O/Fe] abundance ratios of thick-disk stars with $\feh$ up to $\feh\simeq-0.3$ were satisfactorily reproduced using only metallicity-dependent SNII yields in the simple chemical evolution model by R07.

One of the most accepted scenarios for the formation of the thick-disk involves mergers with satellite galaxies early in the history of the Galactic disk \cite[e.g.,][]{quinn93,walker96,velazquez99,abadi03,brook04}. These events could include the following non-exclusive processes: 1) stars from an original disk are heated by interactions with the satellites, which perturb their orbits and increase their eccentricities, 2) stars formed within the merging galaxies are trapped by the potential of the Milky Way's disk as they came close to the plane, and 3) new stars form from freshly accreted gas. At these early times, no significant contribution to the chemical enrichment of the interstellar medium by SNIa occurred, leaving SNII, the end products of massive stars' evolution, as the main polluters. Once settled, the remaining gas from the original disk is flattened by the Galactic potential and disk rotation, from where the present day thin-disk stars are born. This could have happened a few Gyr after the formation of the thick disk. By this time, low-mass stars had sufficient time to evolve, become white dwarfs, and explode as SNIa (if in a binary system, under the right conditions to accrete enough mass, as the most accepted theory for SNIa explosions suggests). The large amounts of iron introduced into the interstellar medium by SNIa quickly decreased the [O/Fe] abundance ratios, explaining the relatively steep decline of [O/Fe] with [Fe/H] for thin-disk stars.

\begin{figure}
\includegraphics[bb=50 355 510 687,width=8.9cm]{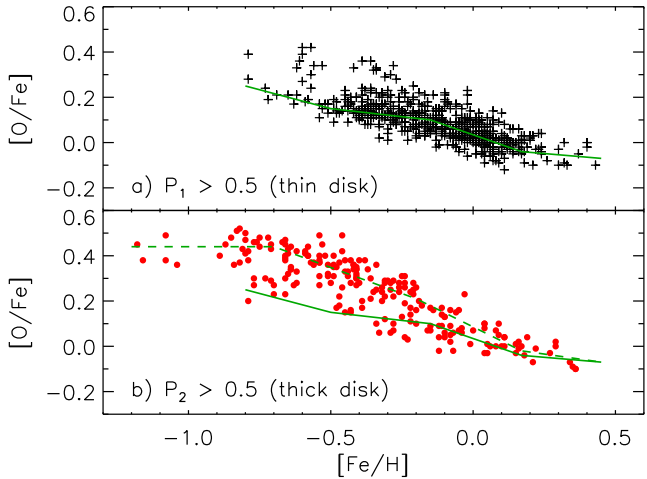}
\caption{{\bf a)} [O/Fe] versus [Fe/H] for kinematically-selected thin-disk stars. The solid line was drawn by hand to trace the behavior of the bulk of these data. {\bf b)} As in a), but for thick-disk stars. The dashed line was drawn by hand to trace the relation followed by most of these data. The solid line from panel a) is overplotted.}
\label{f:ofe_good}
\end{figure}

Our data appear to support the general description given above (Figures~\ref{f:ofe_good}a and b), although the thin/thick disk separation is not as obvious as previously reported, but note that our kinematic membership criterion is not very strong, as we adopt $P_1>0.5$ for thin-disk stars and $P_2>0.5$ for thick-disk stars. As a sample, kinematically-selected thin-disk stars do have lower [O/Fe] ratios than thick-disk stars below $\feh\simeq-0.2$, but a number of these stars have [O/Fe] abundance ratios very similar to those of thick-disk stars of similar metallicity. In fact, below $\feh\simeq-0.5$, the number of (kinematically-selected) thin-disk stars with low [O/Fe] seems to be about the same as that of thin-disk stars with high [O/Fe]. A similar observation can be made regarding kinematically-selected thick-disk stars, i.e., there are a good number of those objects with [O/Fe] abundance ratios very similar to those seen in the mean thin disk. \cite{reddy06} referred to these stars with mixed kinematics and abundances as ``TKTA'' stars.

\begin{figure}
\includegraphics[bb=50 355 510 690,width=8.9cm]{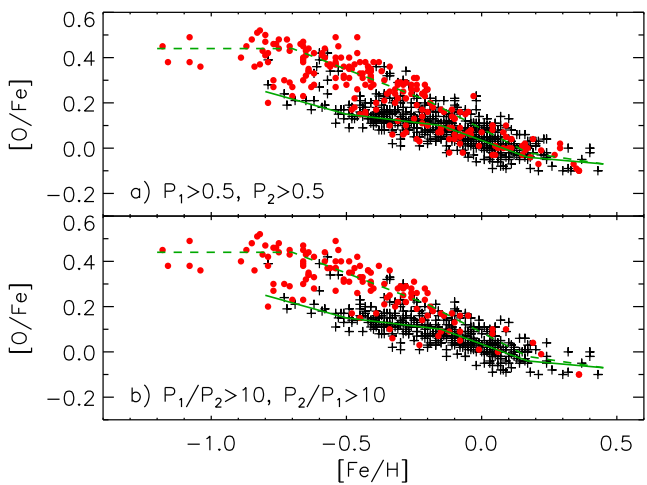}
\caption{{\bf a)} [O/Fe] versus [Fe/H] for kinematically-selected thin-disk stars (crosses) and thick-disk stars (filled circles). {\bf b)} As is a), but for a stronger kinematic membership criterion. The solid and dashed lines are as in Figure~\ref{f:ofe_good}.}
\label{f:ofe_disk}
\end{figure}

Note that in Figure~\ref{f:ofe_good} all of the disk stars from our sample with reliable abundances are plotted. In many previous works regarding elemental abundance differences between thin- and thick-disk stars, stars with kinematics intermediate between thin- and thick-disk stars are excluded from the analysis and chemical evolution interpretation. This is done primarily to prevent kinematically-heated thin-disk stars from ``contaminating'' the thick-disk star sample. In R07, for example, we examined the abundance patterns using only stars with kinematic probabilities greater than 70\,\% (instead of 50\,\% as in this work). Other works have applied an even stronger criterion of $P_1/P_2>10$ for the thin disk (in addition to $P_1>0.5$) and $P_2/P_1>10$ for the thick disk (in addition to $P_2>0.5$). The thin/thick disk oxygen abundance patterns that we obtain using both our soft and the strong kinematic membership criterion are shown in Figure~\ref{f:ofe_disk}. The number of TKTA stars is reduced using the strong criterion, but they do not disappear completely. We find that about 10\,\% of stars with $P_1>0.5$, i.e., kinematic thin-disk stars, have thick disk abundances. Conversely, nearly 20\,\% of kinematically-selected thick-disk stars ($P_2>0.5$) have thin-disk abundances. If we adopt the strong kinematic criterion, these numbers reduce to 8\,\% and 14\,\%, respectively.

The median error in our [O/Fe] measurements is about 0.06\,dex. Assuming a normal distribution for the [O/Fe] errors, we expect only about 1\,\% of thin-disk stars to have thick-disk abundances, and vice versa. In a more simple way, one should realize that if TKTA stars were due to under- or over-estimated [O/Fe] values, one would expect to observe a non-negligible number of thin-disk stars with [O/Fe] significantly lower than the mean [O/Fe] trend of the thin disk, and, similarly, a non-negligible number of thick-disk stars with [O/Fe] significantly larger than the mean [O/Fe] trend of the thick disk. We do not see that in our data, which implies that errors in our abundance analysis can not explain the observed number of TKTA stars.

Of course, the probabilistic approach that we employ must be taken into account when trying to interpret the observed fraction of TKTA stars. We can estimate the expected fraction of stars showing thick-disk (or halo) abundances to which we would erroneously assign thin-disk kinematics, assuming a perfect separation in the [O/Fe] vs.\ [Fe/H] plane, as $n_1^{-1}\sum_{P_1>0.5}(P_2+P_3)$, where $n_1$ is the number of stars with $P_1>0.5$ and the sum extends only to those $n_1$ stars. The fraction of stars with thick-disk kinematics, but thin-disk abundances can be estimated using a similar formula: $n_2^{-1}\sum_{P_2>0.5}(P_1+P_3)$. Interestingly, the expected fractions of TKTA stars are $9.2\pm0.2$\,\% and $19.1\pm0.6$\,\%, in good agreement with the observed fractions of about 10\,\% and 20\,\%, respectively. The error bars reflect the fact that the membership probabilities have uncertainties associated to errors in the $U,V,W$ velocities, which have been propagated into these calculations.

The expected and observed fractions of TKTA stars are very similar, implying that the existence of TKTA stars is a natural consequence of the overlap in the kinematic distributions of thin- and thick-disk stars. However, it should be noted that when a strong kinematic criterion is employed, therefore minimizing the impact of the overlap, the expected fractions of TKTA stars reduce to $2.5\pm0.1$\,\%\ and $11.3\pm0.7$\,\%. These values are smaller than the observed fractions of 8\,\% and 14\,\%, particularly for the case of kinematic thin-disk stars. We note that these fractions are not very sensitive to our particular choice of kinematic parameters for the thin/thick disks. If, for example, instead of using the thick disk velocity dispersions by \cite{soubiran03} we adopt those by \cite{robin03}, which are up to 12\,\kms\ larger, these values remain nearly unchanged; in fact they are different only by 0.1 and 0.3\,\%\ respectively.

We must therefore conclude that it is not possible to tell from the full sample of stars whether the thin- and thick-disk populations are fully separable in kinematics and abundances, because of the overlap in the kinematic distributions, even though the expected and observed fractions of TKTA stars appear to be in good agreement. This is because the inconsistency between the expected and observed fractions of TKTA stars when a strong kinematic separation is made weakens the assumption of duality. By relaxing the condition of perfect thin/thick separation, other possibilities to explain the variety of chemo-dynamical properties of solar neighborhood stars can be explored.

It is possible that stars born in the thin-disk, with thin-disk abundances, have been heated by secular interactions, for example via collisions with molecular clouds or spiral arms, which would naturally explain the existence of kinematic thick-disk stars with thin-disk abundances. The existence of kinematic thin-disk stars with thick-disk abundances, however, cannot be explained by the same types of processes. Perhaps this secular heating is the reason why we have about twice as many thick-disk stars with thin-disk abundances relative to thin-disk stars with thick-disk abundances in our sample (relative to the total number of stars in each population). Another mechanism may mix equally these two populations in kinematics and abundances.

Admittedly, our sample has largely unknown selection functions. It was constructed by collecting high-quality data from a number of sources. Thus, we cannot guarantee that the fractions of TKTA stars quoted before are representative of a volume-limited solar neighborhood sample. Nevertheless, since these objects appear to have been excluded from most previous works, those fractions are likely lower limits to the real frequencies of TKTA stars.

\subsection{Disk Stars with Intermediate Kinematics} \label{s:disk_intermediate}

\begin{figure}
\includegraphics[bb=65 340 510 583,width=8.9cm]{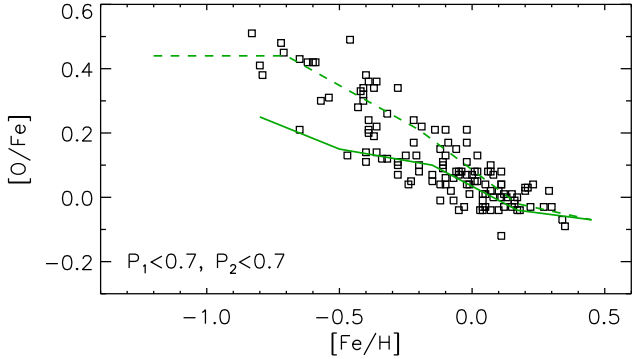}
\caption{[O/Fe] versus [Fe/H] relation for stars with kinematics intermediate between those of thin- and thick-disk stars. The solid and dashed lines are as in Figure~\ref{f:ofe_good}.}
\label{f:ofe_intermediate}
\end{figure}

Stars with kinematics intermediate between those of thin- and thick-disk stars, defined here as non-halo stars ($P_3<0.5$) with $P_1<0.7$ and $P_2<0.7$, appear to have an [Fe/H] distribution slightly more similar to that of the thin disk than the thick disk, i.e., they are, as a sample, more metal-rich than the kinematically warmer stars. However, these ``intermediate kinematic'' (IK) stars do not have a metallicity distribution intermediate between those of the thin- and thick-disk stars. Moreover, their [O/Fe] abundance ratios do not reside in the intermediate region, as shown in Figure~\ref{f:ofe_intermediate}. In fact, these stars seem to cluster around the thin-disk or the thick-disk oxygen abundance trends, but mostly the former, as expected given the larger number of thin-disk stars in any given volume of the solar neighborhood. Only at $\feh\simeq-0.4$ there appears to be a hint that stars with intermediate kinematics have intermediate abundances, but with much larger star-to-star scatter than either thin- or thick-disk stars, which is not expected if the scatter is dominated by observational errors alone. In this region, the star-to-star scatter is about 0.15\,dex while the median [O/Fe] error that we estimate is only about 0.06\,dex.

Although stars with intermediate kinematics do not necessarily have intermediate abundances, their exclusion reduces the number of stars in the region of the [O/Fe]--[Fe/H] plane between the mean thin-disk and thick-disk trends. These objects have been excluded from most previous studies which employed strong kinematic membership criteria, leading to a more clear, albeit artificial, separation between the thin disk and thick disk. If anything, Figure~\ref{f:ofe_intermediate} should be more representative, or at least a less biased representation, of the local disk stars than possibly biased plots such as that presented in Figure~\ref{f:ofe_disk}b. For this work, we made a great observational effort to include as many of these IK objects as possible, in an attempt to reduce the bias in the samples used to investigate the nature of the Galactic disk.

\begin{figure}
\includegraphics[bb=90 370 420 742,width=9.0cm]{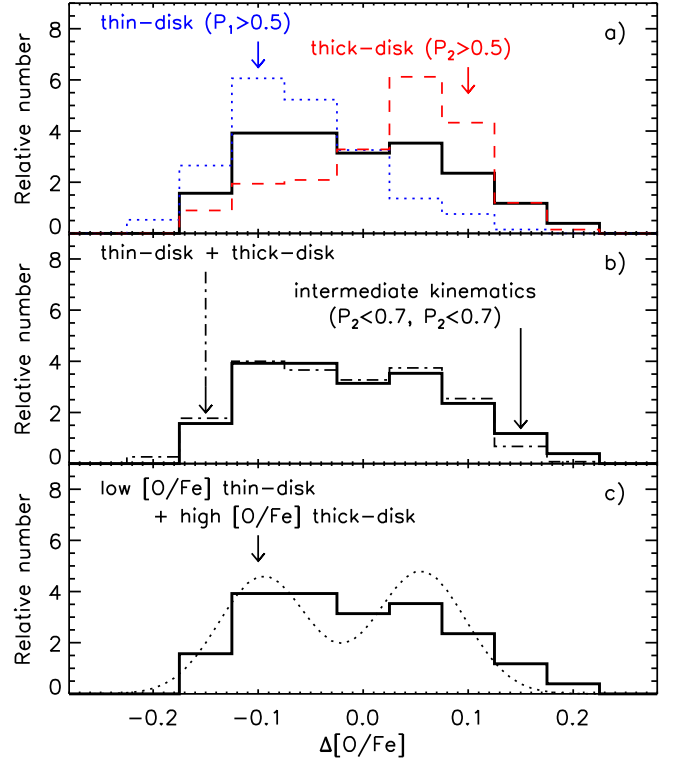}
\caption{{\bf a)} Histogram of $\Delta$[O/Fe] (difference in oxygen-to-iron abundance ratio relative to the line dividing the mean thin- and thick-disk [O/Fe] patterns at $-0.7<\feh<-0.1$) for stars with intermediate kinematics (thick solid line). Similar distributions are shown for thin-disk (dotted line) and thick-disk (dashed line) stars. {\bf b)} The distribution of intermediate kinematics stars is compared to that of the thin- and thick-disk stars as a single group (dot-dashed line). {\bf c)} The distribution of intermediate kinematics stars is compared to that of the combination of low [O/Fe] thin-disk stars and high [O/Fe] thick-disk stars.}
\label{f:disk_histograms}
\end{figure}

A more detailed investigation of the nature of the IK stars can be done using Figure~\ref{f:disk_histograms}, where the distribution of IK stars around the line which divides the average thin- and thick-disk [O/Fe] versus [Fe/H] trends is shown (solid line in Figures~\ref{f:disk_histograms}a, b, and c) along with those of thin-disk ($P_1>0.5$, dotted line in Figure~\ref{f:disk_histograms}a) and thick-disk ($P_2>0.5$, dashed line in Figure~\ref{f:disk_histograms}a) stars. The samples are restricted to the metallicity range $-0.7<\feh<-0.1$, i.e., away from the high metallicity end where the thin- and thick-disk trends overlap, and also away from the low metallicity end which contains few thin-disk stars. The thin-disk and thick-disk peaks are clearly identified on the low $\Delta$[O/Fe] and high $\Delta$[O/Fe] sides, respectively. Their distributions, however, have extended tails towards the intermediate region, a property that is probably due to the inclusion of a significant number of IK stars in each group. The tails of these distributions towards lower $\Delta$[O/Fe] for thin-disk stars and towards higher $\Delta$[O/Fe] for thick-disk stars, on the other hand, are fully consistent with Gaussian tails dominated by the observational errors.

In Figure~\ref{f:disk_histograms}b the distribution of IK stars (solid line) is compared to that of all disk stars in the $\feh$ range mentioned above (dot-dashed line). They are nearly identical, suggesting that the IK sample is indeed a better representation of the Galactic disk as a whole, i.e., a sample of disk stars without strong dual kinematic bias selection. On the other hand, if one neglects the extended tails of the thin- and thick-disk distributions towards higher and lower $\Delta$[O/Fe], respectively, the distribution of disk stars appears clearly double peaked, as shown in Figure~\ref{f:disk_histograms}c (dotted line). This analytical distribution was derived by fitting Gaussian tails to the low [O/Fe] thin-disk and high [O/Fe] thick-disk data separately, and extending those Gaussian distributions towards the intermediate region.\footnote{The dotted line in Figure~\ref{f:disk_histograms}c corresponds to a hypothetical distribution of disk stars obtained using a sample selection function similar to the one employed in this work, but biasing it further towards the extremes of the thin- and thick-disk [O/Fe] distributions. If our sample were volume-limited, we would expect the peak of the ``thin-disk side'' of that distribution ($\Delta\ofe<0$) to be significantly higher than that at $\Delta\ofe>0$, resulting in a Gaussian-like distribution with a single peak at $\Delta\ofe\simeq-0.1$ and a weak $\Delta\ofe>0$ tail. The dotted line in Figure~\ref{f:disk_histograms}c is not meant to represent a volume-limited sample, but a group of stars chosen as in our work.} The IK star sample does not look like this heavily biased group of disk stars at all, i.e., their $\Delta$[O/Fe] distribution is not bimodal. Thus, as we have argued before, IK stars do not populate the region intermediate between those populated by the majority of thin- and thick-disk stars separately in the [O/Fe] versus [Fe/H] plane, but they do contribute more to the frequency of disk stars in that region relative to the case when a highly biased sample of disk stars, in terms of their kinematics, is used.

\subsection{The Thin/Thick Disk Duality} \label{s:disk_duality}

Having established that stars with ``ambiguous'' kinematics and abundances, i.e., TKTA stars as well as stars with intermediate [O/Fe] abundance ratios, are not uncommon in the solar neighborhood, one must question whether models predicting (or assuming) a thin/thick disk duality are realistic. We should begin to consider alternative scenarios in which these objects are a natural consequence of the Galaxy's evolution. In this context, it is relevant to mention the importance of radial migration for the formation of galactic disks \cite[e.g.,][]{sellwood02}.

Stars in the solar neighborhood could have been born at different Galactocentric distances, each having a different chemical enrichment history, and brought to the solar vicinity after several billion years of Galactic evolution. Present-day elemental abundance gradients with Galactocentric distance for both oxygen and iron have been observed \cite[e.g.,][]{deharveng00,friel02,jacobson07,henry10}, implying that radial migration provides a way to explain the range of elemental abundances seen in the solar neighborhood. The chemical evolution model by \cite{schonrich09a,schonrich09}, for example, predicts an [O/Fe]--[Fe/H] distribution that is remarkably similar to that seen in our data (compare, for example, their Figure~8 with our Figure~\ref{f:ofe_disk}). Their models ``include radial migration of stars and flow of gas through the disk.'' Our data provide support to the idea that this mechanism is one of the most important ones for shaping the chemo-dynamical properties of stars in the solar neighborhood.

Admittedly, some cosmological simulations \citep[e.g.,][]{kobayashi11} can probably reproduce the observational properties of the solar neighborhood equally well. A very important exercise in this case is, however, to establish from first principles which mechanism(s) is(are) the most relevant one(s) with regards to those properties, and whether such models can produce galaxies like the Milky Way naturally rather than as one particular example in a multitude of test cases.

As argued before, excluding stars with kinematics intermediate between thin- and thick-disk stars does not affect in a dramatic manner the observed trends, since the intermediate kinematics stars do not {\it all} define a clear intermediate abundance pattern, but rather a majority of them tend to follow those of either the thin disk or the thick disk. Nevertheless, this does not imply that the thin and thick disk abundance patterns are {\it fully} separable in both elemental abundances and kinematics. As larger samples have been assembled, and very strong kinematic membership criteria are not applied, the thin and thick disk oxygen abundance trends have began to become less well defined thanks to the larger number of TKTA and IK stars analyzed. Our nearby star data do not fully support the idea of a dual thin/thick disk. Interestingly, analyses of [$\alpha$/Fe] data from SEGUE/SDSS by \cite{bovy12_vertical,bovy12_spatial} suggest that the disk chemo-dynamical properties could be well fitted as a single population, questioning the reality of the thick disk itself \citep{bovy12_nothick}.

\subsection{Two Halo Populations} \label{s:halo}

\begin{figure}
\includegraphics[bb=50 355 510 745,width=8.9cm]{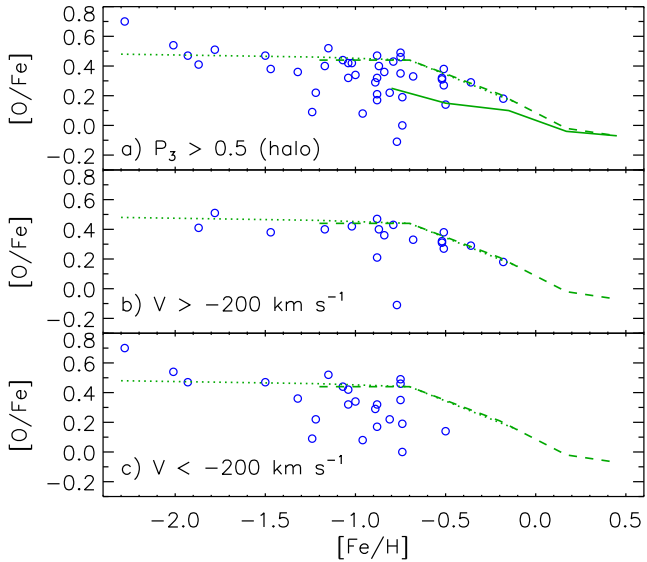}
\caption{{\bf a)} [O/Fe] versus [Fe/H] for kinematically selected halo stars with $V>-200$\,\kms\ {\bf b)} As in a) for halo stars with $V<-200$\,\kms.}
\label{f:ofe_halo}
\end{figure}

Even though the number of halo stars in our sample is relatively small, we can still obtain important information from their oxygen abundance patterns. Figure~\ref{f:ofe_halo}a shows the [O/Fe] versus [Fe/H] relation of halo stars. There is significant star-to-star scatter, much larger than that seen in the disk stars. Although the abundance errors are larger for this sample of more distant, fainter stars with significantly weaker spectral lines relative to the disk sample, they cannot fully explain the large scatter observed here. Interestingly, we find that most of the scatter is due to stars with a Galactic space velocity $V<-200$\,\kms, as shown in Figures~\ref{f:ofe_halo}b and c.

\cite{nissen10} have shown that in the disk/halo transition region of the $[\alpha/\mathrm{Fe}]$ versus $\feh$ plane, i.e., at about $-1.5<\feh<-0.5$, stars with halo kinematics separate into well-defined samples of low-$\alpha$ and high-$\alpha$ stars, based on their $[\alpha/\mathrm{Fe}]$ abundance ratios, where the $\alpha$ elements in their study are Mg, Si, Ca, and Ti. They also find that stars with thick-disk kinematics in the same metallicity range follow almost exactly the trend defined by the high-$\alpha$ halo stars. Low- and high-$\alpha$ halo stars have also separate Cu, Zn, and Ba to Fe abundance ratio trends \citep{nissen11}, but there is no significant difference regarding their lithium abundances \citep{nissen12}. Age and kinematic information useful to investigate the nature of these halo sub-populations has been provided by \cite{schuster12}. Oxygen abundances for the majority of stars in the \cite{nissen10} study have been recently derived by \cite{ramirez12_halo2pop}, who find that the low- and high-$\alpha$ halo stars also separate in the [O/Fe] versus [Fe/H] plane and that thick-disk stars and high-$\alpha$ halo stars follow nearly indistinguishable oxygen abundance patterns.

As shown by \cite{nissen10}, an important fraction of the low-$\alpha$ halo stars are retrograde regarding their Galactic rotation velocities. In Figures~\ref{f:ofe_halo}b and c we separate our halo stars in two groups of $V>-200$\,\kms\ and $V<-200$\,\kms. This allows to separate the majority of stars in retrograde orbits from the rest of halo stars. Clearly, most of the scatter seen in our [O/Fe] versus $\feh$ relation comes from stars in retrograde orbits. Note, however, that not all stars with retrograde orbits are low [O/Fe] stars, an observation that is also consistent with the work by \cite{nissen10}. The star with low $\mathrm{[O/Fe]}\simeq-0.1$ in Figure~\ref{f:ofe_halo}b is HIP\,12294, which has uncertain astrometric data, and therefore a $V$ velocity with a large error bar ($V=-154.5\pm52.4$\,\kms). Excluding it from that plot, we find that the high-$\alpha$, ``high-oxygen'' halo stars from our sample follow a tight [O/Fe] versus [Fe/H] relation such that $\ofe$ is nearly constant at $\ofe\simeq0.45$ from $\feh\simeq-2.0$ to $\feh\simeq-0.7$, from where it decreases smoothly down to $\ofe\simeq0.2$ at $\feh\simeq-0.2$. The star-to-star scatter of this relation is 0.065\,dex, which is only slightly larger than the dispersion predicted by our observational uncertainties, leaving little or no room for cosmic scatter within this group of stars.

Thus, our oxygen abundance results for halo stars support the observations by \cite{nissen10} regarding $\alpha$-element abundances, and therefore also suggest that there is a certain degree of heterogeneity in the abundances and kinematics of nearby halo stars. \cite{nissen10} argue that the halo is composed of two discrete populations, and that perhaps the halo stars with low $\alpha$-element abundances (and therefore those with low [O/Fe] abundance ratios in our work) could have been accreted from dwarf satellite galaxies, most notably the present-day globular cluster $\omega$\,Cen, whereas the high-$\alpha$ stars were born within the Galaxy \cite[see also][]{nissen11,schuster12}. Note, however, that the $\omega$\,Cen connection is questionable considering the complexity of that particular globular cluster whereas \cite{ramirez12_halo2pop} argue that at least two of the low-$\alpha$ field halo stars from the \cite{nissen10} sample were likely born in globular clusters, a claim based on the very low oxygen abundances and very high sodium abundances of those two objects.

A dual scenario seems to explain our nearby star halo data. As the halo was forming, stars from the Galaxy's building blocks, some of which may be the present-day dwarf satellites, were captured by the early Milky Way into highly eccentric orbits, including retrograde orbits. The chemical composition of these objects reflects that of their parents, i.e., systems with slow star formation rate which lead to low [O/Fe] as well as low [$\alpha$/Fe] abundance ratios at relatively low $\feh$ because the contribution of SNII is reduced. Observations of halo streams and tidal debris heavily support the idea of early mergers \cite[e.g.,][]{helmi08,klement10,majewski12}. Gas belonging to the early halo itself, with more massive stars forming and exploding as SNII, was used to form the present-day high-$\alpha$, high-oxygen halo stars. Since this gas was probably much more homogeneous than that of the mixture of dwarf satellite galaxies, this scenario predicts a small star-to-star scatter in the abundance ratios of their surviving members.

Other recent works point to a halo dichotomy similar to that described above. Using data from SDSS, for example, \cite{carollo07,carollo10} find evidence for an outer halo that has a net retrograde rotation, contrary to the inner halo, which is, in addition, slightly more metal-rich \cite[see also][]{beers12}. This result has been questioned by \cite{schonrich11} on the basis of a possibly biased distance determination \cite[but see][]{beers12}. On the other hand, using also SDSS data, \cite{jofre11} find that main-sequence turn-off (MSTO) stars in the halo consist of a dominant population that formed quickly, but they find evidence for a significant number of MSTO halo stars younger than the dominant population, and argue that the latter may be the remnants of early accretion of external galaxies. Interestingly, \cite{schuster12} find that the low-$\alpha$ halo stars are about 2--3\,Gyr younger than the high-$\alpha$ members, supporting this idea.

\subsection{Chemical Tagging in the Galactic Disk}

\begin{figure}
\includegraphics[bb=68 370 450 885,width=8.95cm]{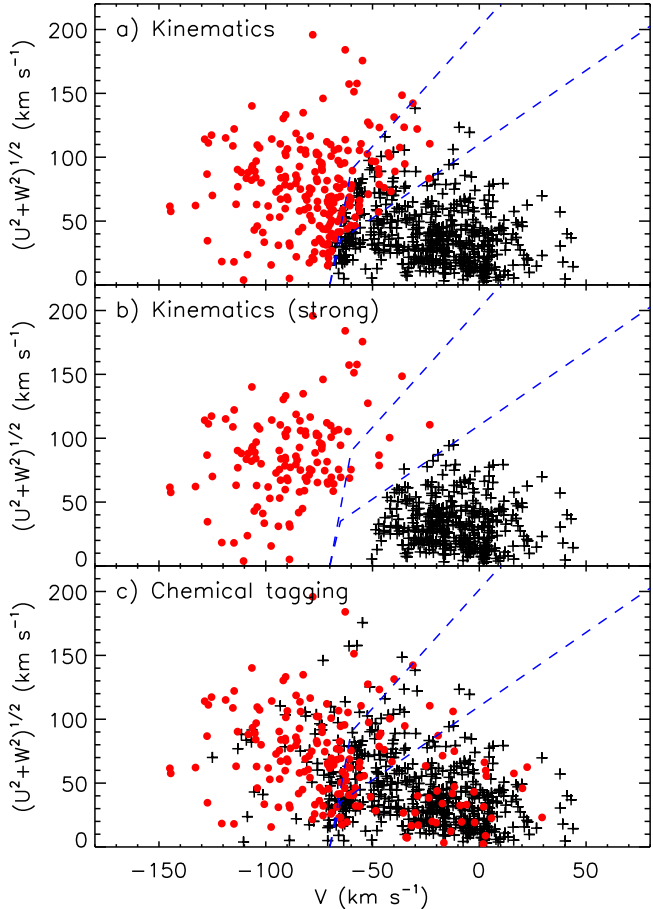}
\caption{{\bf a)} Toomre diagram of stars with $P_1>0.5$ (crosses) and $P_2>0.5$ (filled circles). {\bf b)} Toomre diagram of stars with $P_1/P_2>10$ (crosses) and $P_2/P_1>10$ (filled circles). {\bf c)} Toomre diagram of stars separated using chemical tagging, as shown in Figure~\ref{f:ofe_chemtag}. The dashed lines represent the kinematic thin/thick disk boundary.}
\label{f:toomre_chemtag}
\end{figure}

Instead of using kinematics to separate the stars into thin- and thick-disk objects, as described in previous Sections, and as illustrated by Figures~\ref{f:toomre_chemtag}a and b, one could use the elemental abundances to identify the stars as members of one of these two groups. In general, this approach is referred to as chemical tagging. The difference between panels a and b in Figure~\ref{f:toomre_chemtag} is that the latter (panel b) corresponds to the strong kinematic membership criterion (i.e., $P_1/P_2>10$ for the thin disk, $P_2/P_1>10$ for the thick disk), and therefore excludes a significant number of stars from the analysis.

\begin{figure}
\includegraphics[bb=70 370 510 625,width=8.9cm]{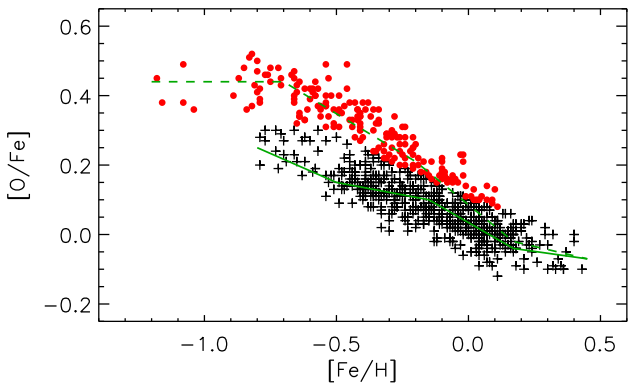}
\caption{Disk stars are separated into two groups using their [O/Fe] abundance ratios. The boundary is somewhat arbitrary, but it has been set so that the mean trends of these groups roughly correspond to those of kinematically-selected thin- and thick-disk stars. Solid and dashed lines are as in Figure~\ref{f:ofe_disk}.}
\label{f:ofe_chemtag}
\end{figure}

In Figure~\ref{f:ofe_chemtag}, Galactic disk stars have been divided into two groups according to their [O/Fe] abundance ratios for any given [Fe/H]. The separation was defined in such a way that, on average, the two groups resemble the trends followed by the bulk of kinematically-selected thin- and thick-disk star samples. A Toomre diagram showing the location of these two groups of stars is plotted in Figure~\ref{f:toomre_chemtag}c.

Obviously, chemical tagging of Galactic disk stars does not result in a perfect kinematic separation, although average tendencies are detected such that low-[O/Fe] stars tend to have smaller $|V|$ and $(U^2+W^2)^{1/2}$ velocities. High-[O/Fe] stars, on the other hand, rotate slower around the center of the Galaxy (i.e., have more negative $V$ velocities) and their orbits reach higher altitudes from the Galactic plane. However, we must not neglect the fact that there are many low-[O/Fe] stars with warm kinematics, as well as many kinematically cold high-[O/Fe] stars, reminiscent of TKTA stars.

The dashed lines in Figure~\ref{f:toomre_chemtag} roughly represent the boundary between kinematically-selected thin- and thick-disk stars. The region contained between the two dashed lines has stars of both populations, but the regions outside of this area are unique to members of one group or the other. By counting the number of stars on the left and right hand sides of this boundary, we can determine the fraction of stars with low-[O/Fe] (high-[O/Fe]) and warm (cold) kinematics. We find those numbers to be 16\,\% and 13\,\%, respectively. Observational errors alone allow these values to be only about 2\,\%.

Both the kinematic and chemical tagging methods of separating disk stars into two groups end up with a significant number of stars with ambiguous kinematics and abundances, a number that is difficult to explain with our estimates of the observational errors. A simple picture of kinematically-selected thin-disk stars with low [O/Fe] and thick-disk stars with high [O/Fe] seems to no longer apply.

\subsection{Trends with Age}

\begin{figure}
\includegraphics[bb=60 365 450 885,width=8.9cm]{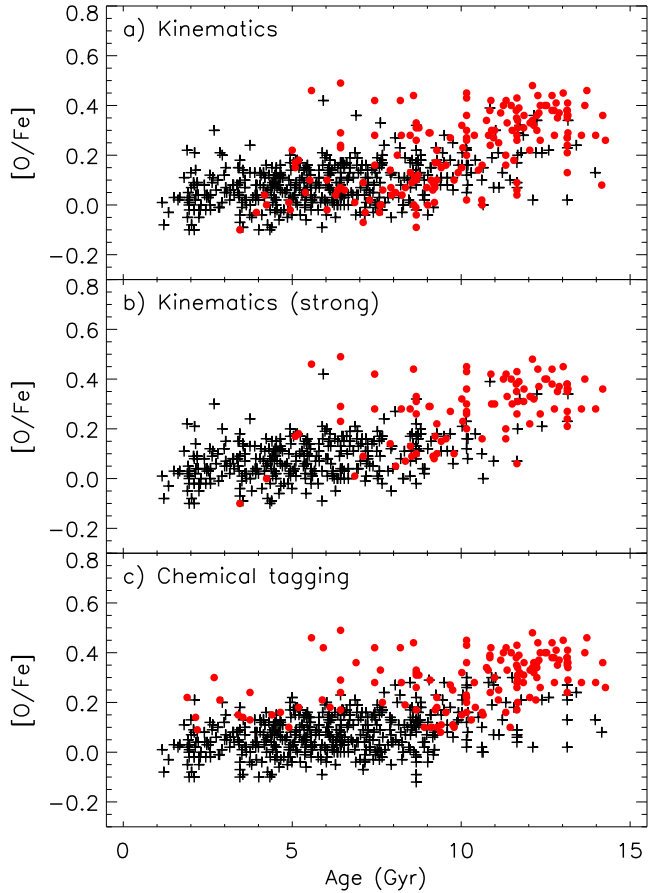}
\caption{{\bf a)} [O/Fe] versus age for disk stars separated using their kinematics ($P_1>0.5$ for the thin disk and $P_2>0.5$ for the thick disk). {\bf b)} As in panel a) but for a strong kinematic criterion ($P_1/P_2>10$ for the thin disk and $P_2/P_1>10$ for the thick disk). {\bf c)} [O/Fe] versus age for disk stars separated using chemical tagging, as in Figure~\ref{f:ofe_chemtag}.}
\label{f:ofe_age}
\end{figure}

The age distributions of thin- and thick-disk stars, selected using kinematics, are shown in Figures~\ref{f:ofe_age}a and b. The latter corresponds to the case of strong kinematic separation. In general, kinematically-selected thick-disk stars are older than thin-disk members. However, there is considerable overlap between the two groups when all stars are plotted, and some overlap when a strong kinematic separation criterion is applied. In Figure~\ref{f:ofe_age} (and hereafter), only our sample stars with ages greater than their 3\,$\sigma$ error are used.

Kinematically-selected thin- and thick-disk stars do have different age distributions, with the former being younger than the latter, but we do not find in our data convincing evidence for a sharp boundary between the two groups. When a strong kinematic criterion is adopted, the analysis of a relatively small sample of disk stars may lead one to conclude that there are no kinematically-selected thick-disk stars below a certain age. Note that there are only a handful of thick disk stars younger than about 6-8\,Gyr in Figure~\ref{f:ofe_age}b, which, due to their younger ages and the fact that they are mainly solar-type stars, are probably close to the main-sequence and have relatively large age errors. Thus, these stars could have been excluded in previous works because of that, leading to a sharp lower limit for the age distribution of thick-disk stars, which may be, however, spurious.

In Figure~\ref{f:ofe_age}c, the age distributions of stars separated according to their [O/Fe] abundance ratios, as in Figure~\ref{f:ofe_chemtag}, are shown. Very few stars with $\ofe>0.2$ and younger than 5\,Gyr are found. Although the majority of high [O/Fe] stars are older than about 10\,Gyr, a number of these stars are found with ages in the range from 5 to 10\,Gyr. Also, note the presence of a few stars with low $\ofe\simeq+0.1$ and very old ages.

An overall trend of increasing [O/Fe] with older age is clearly observed in Figure~\ref{f:ofe_age}, but once again it does not separate perfectly thin and thick disk stars, regardless of how these groups are defined, i.e., using their kinematics or with chemical tagging. There is also significant scatter in the [O/Fe] versus age relation, which could in part be due to observational errors, but the presence of a non-negligible number of thin-disk stars with very old ages as well as young kinematic thick-disk stars suggests that a sharp thin/thick disk separation is unrealistic.

\begin{figure}
\includegraphics[bb=55 365 480 705,width=8.65cm]{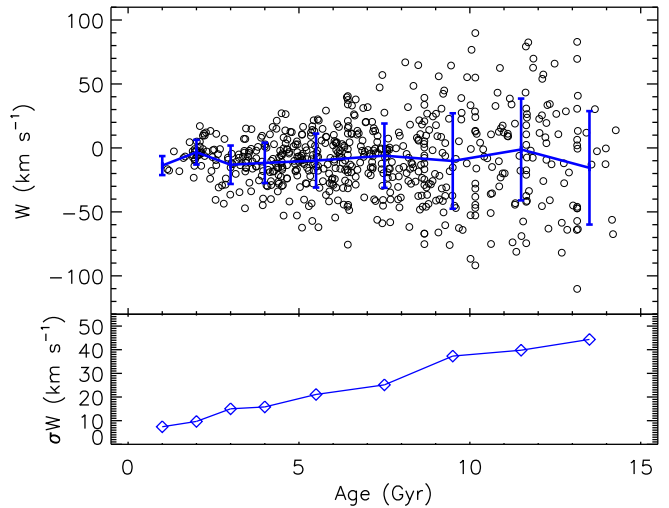}
\caption{Top panel: Galactic space $W$ velocities as a function of age for all disk stars. The solid line with error bars represent the average value and star-to-star scatter (1\,$\sigma$) at a given age. Bottom panel: Star-to-star scatter in $W$ (i.e., $\sigma W$) as a function of age for all disk stars.}
\label{f:w_age}
\end{figure}

\begin{figure}
\includegraphics[bb=75 355 480 575,width=8.9cm]{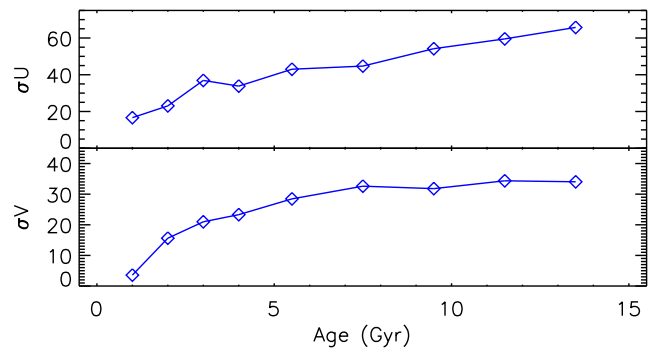}
\caption{Star-to-star scatter in $U$ (top panel) and $V$ (bottom panel) as a function of age for all disk stars.}
\label{f:uv_age}
\end{figure}

The velocity dispersion of disk stars increases with age, as shown for the $W$ component in Figure~\ref{f:w_age}. This result, which is based on all disk stars with reliable ages, is in excellent agreement with that found by, e.g., the GCS \citep{nordstrom04}. The $1\,\sigma$ star-to-star scatter in the $W$ values of disk stars increases from about 10\,\kms\ at 1\,Gyr to 40\,\kms\ at 10\,Gyr and older ages, as shown in the bottom panel of Figure~\ref{f:w_age}. The $U$ and $V$ velocity dispersions of disk stars in our sample also increase with age, as shown in Figure~\ref{f:uv_age}.

\begin{figure}
\includegraphics[bb=75 370 480 745,width=9.0cm]{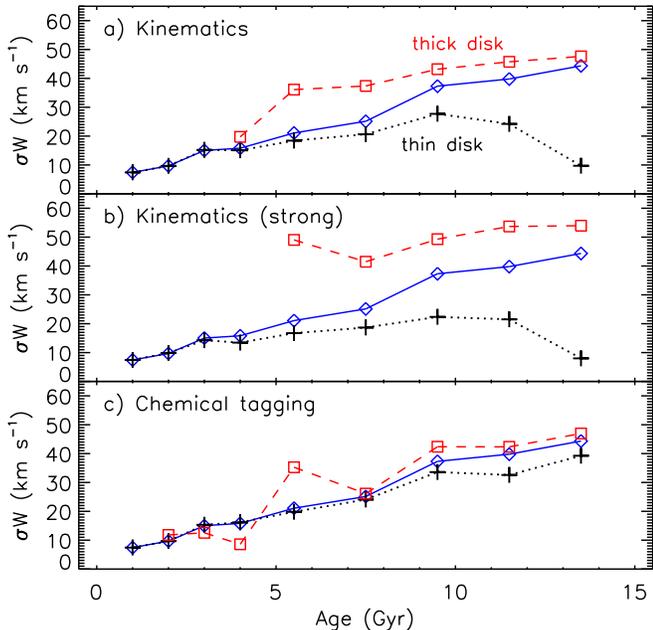}
\caption{{\bf a)} The star-to-star scatter in $W$ as a function of age for all disk stars (diamonds and solid line) is compared to that of kinematically-selected thin- (crosses and dotted line) and thick-disk (squares and dashed line) stars. {\bf b)} As in panel a) but with a strong kinematic selection. {\bf c)} As in panel a) but employing chemical tagging.}
\label{f:sw_age}
\end{figure}

If the stars are separated as thin- and thick-disk members, one could conclude that the increase in the $W$ velocity scatter is due to thick-disk stars only. In Figure~\ref{f:sw_age}a we note that $\sigma W$ increases only slightly with age for kinematically-selected thin-disk stars, and much more rapidly, as well as at a higher level, for thick-disk objects. On the other hand, if the thin/thick disk membership is assigned using the strong kinematic criterion, the $\sigma W$ of thick-disk stars appears to be nearly constant with age, at about 50\,\kms, as shown in Figure~\ref{f:sw_age}b, while the thin-disk $\sigma W$ values do not exceed 20\,\kms\ at any given age.

The results described above suggest that a proper kinematic membership assignment must take into account the fact that the velocity dispersions of the two populations depend on age. Although the velocity dispersions as a function of age could be determined with reasonable precision using large survey data, the uncertainties in the determination of ages for individual stars prevent us from adopting this approach. Moreover, for many stars no meaningful age can be derived, implying that they will be excluded from the analysis, potentially introducing another sample bias.

When disk stars are separated according to their [O/Fe] abundance ratios, the run of $\sigma W$ with age is very similar for both low [O/Fe] (``thin-disk'') stars and high [O/Fe] (``thick-disk'') stars, as shown in Figure~\ref{f:sw_age}c. The $\sigma W$ values are less certain for the thick-disk at young ages due to the low number of young thick-disk stars used to compute the velocity dispersion. This explains the ``noisy'' nature of the $\sigma W$ versus age relation of chemically-tagged thick-disk stars at young ages (the same argument can be made to explain the peculiarly low $\sigma W$ value for the oldest kinematically-selected thin-disk stars in Figures~\ref{f:sw_age}a and b). Admittedly, Figure~\ref{f:sw_age}c suggests that thick-disk stars still have slightly higher $\sigma W$ at any given age relative to thin-disk stars, but the differences are not as dramatic as seen in the former case, i.e., when the stars' kinematic properties are used to define these two samples. That the latter happens is not at all surprising, and one could argue that this is a redundant statement, but the fact that chemically-tagged thin- and thick-disk stars show an increase in velocity dispersion with age nearly identical suggests again that they may be in fact members of a single stellar population.

\section{FINAL REMARKS} \label{s:conclusions}

We have measured oxygen abundances of a large sample of nearby stars, mostly from the Galactic disk, but including also a number of halo members. Significant improvements to the stellar parameter determination methods have been made with respect to our previous work on this topic. Moreover, we have increased the number of stars in our study by almost 60\,\%. In particular, we now include an important number of objects with kinematics intermediate between those typically assigned to the thin disk and thick disk sub-samples. The inclusion of these objects allows us to investigate the elemental abundance patterns in a less biased way. Most previous works have completely ignored these objects in an attempt to avoid sample ``contamination,'' but we argue that such approach may have led to spurious conclusions.

The simple picture of an old, kinematically warm thick disk composed of stars with [O/Fe] abundance ratios that are always greater than those of thin-disk stars at $\feh\lesssim-0.1$, which are younger and kinematically cold as a sample, is challenged by our data. We find significant overlap in the chemical distributions of kinematically-selected thin- and thick-disk stars. Observational errors can explain a fraction of kinematically-selected thin-disk stars with thick-disk abundances, and vice-versa, but the fractions of these stars (which we refer to as TKTA stars) observed in our data are higher than expected on the basis of random errors alone. Although the observed fractions of TKTA stars are compatible with expectations when a {\it weak} kinematic selection criterion is used (a star is considered to be part of a population when its membership probability $P_i$ is higher than 0.5), this is no longer true when a {\it strong} kinematic criterion is adopted (i.e., $P_1/P_2>10$, $P_2/P_1>10$). Thus, alternatives to the dual thin/thick disk picture need to be explored.

We find that stars with thin-disk abundances but thick-disk kinematics outnumber by about a factor of two the stars with thick-disk abundances but thin-disk kinematics. We attribute this observation to secular perturbations of ``true'' thin-disk star orbits, which, however, do not explain the thick-disk counterparts. We have looked in detail at stars with kinematics intermediate between those of the majority of thin- and thick-disk stars, which are probably more numerous (in relative numbers) in our work than in previous chemical abundance studies of the local Galactic disk. These intermediate kinematics (IK) disk objects do not have an [O/Fe] versus [Fe/H] relation intermediate between those of the average thin- and thick-disk stars. Nevertheless, they populate the intermediate region better than any of those two groups separately, but almost equally if a sample of kinematically-selected thin-disk stars is combined with a similarly chosen sample of thick-disk objects. This implies that our IK sample is more representative of the local Galactic disk as a whole than the heavily biased samples of thin- or thick-disk stars selected using strong kinematic criteria.

The large size of our sample and our new observations of an important number of IK stars have allowed us to see a less biased representation of the chemical properties of the local Galactic disk, albeit using only the very important element oxygen. The chemical evolution of the solar neighborhood appears more complex than previously thought. With the analysis of high-quality spectroscopic data for larger, volume-limited samples, or extended samples of stars with very well known selection functions, we will be able to better trace the history of formation and evolution of the Galactic disk, the reality of which is, in fact, not too far ahead into the future.

\acknowledgments

I.R.'s work was performed under contract with the California Institute of Technology (Caltech) funded by NASA through the Sagan Fellowship Program. D.L.L. acknowledges support from the Robert A. Welch Foundation of Houston, Texas (grant number F-634). The authors thank the referee for reviewing this paper and providing suggestions to improve it. Some of the data used in this paper are from the UVES Paranal Observatory Project (ESO DDT Program ID 266.D-5655) and the Hobby-Eberly Telescope (HET), which is a joint project of the University of Texas at Austin, the Pennsylvania State University, Stanford University, Ludwig-Maximilians-Universit\"at M\"unchen, and Georg-August-Universit\"at G\"ottingen. The HET is named in honor of its principal benefactors, William~P.\ Hobby, Jr.\ and Robert~E.\ Eberly.

\end{document}